\documentclass[12pt]{article}
\usepackage[latin1]{inputenc}
\usepackage{amssymb}
\usepackage{hyperref}
\usepackage{amsmath}
\usepackage{latexsym}
\usepackage{times}
\usepackage{cite}
\usepackage{amssymb}
\usepackage{amsfonts}
\usepackage{amscd}
\usepackage{xcolor}
\usepackage{amstext,amsmath,amssymb,amsfonts}
\newtheorem{definition}{Definition}

\newtheorem{theorem}[definition]{Theorem}

\newtheorem{remark}[definition]{Remark}

\newtheorem{proposition}[definition]{Proposition}

\newcommand{\n}{\mathcal{N}}

\newcommand{\beq}{\begin{eqnarray}}
\newcommand{\eeq}{\end{eqnarray}}
\newcommand{\beqs}{\begin{eqnarray*}}
	\newcommand{\eeqs}{\end{eqnarray*}}
\newcommand{\bpro}{\begin{pro}}
	\newcommand{\epro}{\end{pro}}
\newcommand{\blem}{\begin{lem}}
	\newcommand{\elem}{\end{lem}}
\newcommand{\bdfn}{\begin{dfn}}
	\newcommand{\edfn}{\end{dfn}}
\newcommand{\bcor}{\begin{cor}}
	\newcommand{\ecor}{\end{cor}}
\newcommand{\bthm}{\begin{thm}}
	\newcommand{\ethm}{\end{thm}}
\newcommand{\bex}{\begin{ex}}
	\newcommand{\eex}{\end{ex}}
\newcommand{\brmk}{\begin{rmk}}
	\newcommand{\ermk}{\end{rmk}}
\newcommand{\bpr}{\begin{pr}}
	\newcommand{\epr}{\end{pr}}
\newcommand{\benum}{\begin{enumerate}}
	\newcommand{\eenum}{\end{enumerate}}
\newcommand{\bitem}{\begin{itemize}}
	\newcommand{\eitem}{\end{itemize}}

\newcommand{\cqfd}{\hfill{\square}}
\chardef\bslash=`\\

\begin{document}	
	\begin{center}
		{\Large {Multi-parameter  Fermi-Dirac and Bose-Einstein Stochastic Distributions}}\\
		\vspace{0,5cm}
		  Fridolin Melong and  Mahouton Norbert Hounkonnou\\
		\vspace{0.5cm}
		{\em International Chair in Mathematical Physics
			and Applications}
		{\em (ICMPA-UNESCO Chair), }
		{\em University of Abomey-Calavi,}
		{\em 072 B.P. 50 Cotonou,   Benin Republic,} 	\\
		{\em  and  International Centre for Research and Advanced Studies in Mathematical and Computer Sciences, and Applications (ICRASMCSA), 072 B.P. 50 Cotonou, Benin Republic}
		\\
		fridomelong@gmail.com\\
		norbert.hounkonnou@cipma.uac.bj, (with copy to hounkonnou@yahoo.fr)
		
	\end{center}
	\today
	
	\vspace{0.5 cm}
	{\it Abstract}
	In this paper, we characterize  the multivariate uniform probability distribution of the first and second kinds in the framework of the  $\mathcal{R}(p,q)$-deformed quantum
	algebras. Their bivariate distributions and related properties, namely ($\mathcal{R}(p,q)$-mean, $\mathcal{R}(p,q)$-variance and $\mathcal{R}(p,q)$-covariance) are computed and discussed.  
	Particular cases corresponding to quantum algebras
	existing in literature are deduced.

\tableofcontents
	\section{Introduction}
	The generalization of the discrete uniform distribution from the quantum algebra with one parameter (also called $q$-uniform distribution) first emerged as a congruence class distribution,
	modulo $n,$ of Bernoulli generated numbers, in a probabilistic number theory 
	\cite{Rawlings}. A discrete $q$-uniform distribution starting
	with a nonnegative $q$-function defined on the set $\{0, 1, \ldots , n\}$ was discussed by Kupershmidt \cite{Kupershmidt}. Besides, 
	the characterization,  properties, and applications  of the univariate $q$-uniform distribution were presented and investigated by Charalambides \cite{Charalambides2016}. 
	The most important multivariate discrete uniform distributions are interpreted as  the
	Fermi-Dirac and Bose-Einstein stochastic models, with the probability functions \cite{Feller1968}:
	\begin{eqnarray*}
	P\big(X_1=x_1,\ldots,X_k=x_k\big)=1\bigg/{k+1\choose n},
	\end{eqnarray*}
	where $x_j\in\{0,1\},\,j\in\{1,2,\ldots,k\},$  $\sum_{j=1}^{k}x_j\leq n$ and 
	\begin{eqnarray*}
		P\big(X_1=x_1,\ldots,X_k=x_k\big)=1\bigg/{k+n\choose n},
	\end{eqnarray*}
	where $x_j\in\{0,1,\ldots,n\},\,j\in\{1,2,\ldots,k\},$  $\sum_{j=1}^{k}x_j\leq n,$ respectively.

It is worth mentioning that in
a quite close analogy, in combinatorics, Chung and Kang introduced the notion of
a $q$-selection of an element from the set $\mathcal{C}=\{c_1,c_2,\ldots,c_r\}$ by considering a weight $q^{i-1}$ as the payment for $i-1$ jumps made in traveling from the left to the right of 
the permutation $p_r = (c_1, c_2, \ldots, c_r ),$ with $c_1 < c_2 < \ldots < c_r,$ before selecting the
element $c_i\in \mathcal{C}$ \cite{ChungKang}.

Let us introduce the notions of  $\mathcal{R}(p,q)$-calculus (differentiation and integration) and gneralized quantum algebras.
	For that,
	let $p$ and $q$ be two positive real numbers such that $ 0<q<p<1.$ We consider a meromorphic function ${\mathcal R}$ defined on $\mathbb{C}\times\mathbb{C}$ by\cite{HB}:\begin{equation}\label{r10}
	\mathcal{R}(u,v)= \sum_{s,t=-l}^{\infty}r_{st}u^sv^t,
	\end{equation}
	with an eventual isolated singularity at the zero, 
	where $r_{st}$ are complex numbers, $l\in\mathbb{N}\cup\left\lbrace 0\right\rbrace,$ $\mathcal{R}(p^n,q^n)>0,  \forall n\in\mathbb{N},$ and $\mathcal{R}(1,1)=0$ by definition. We denote by $\mathbb{D}_{R}$ the bidisk \begin{eqnarray*}
		\mathbb{D}_{R}
		=\left\lbrace w=(w_1,w_2)\in\mathbb{C}^2: |w_j|<R_{j} \right\rbrace,
	\end{eqnarray*}
	where $R$ is the convergence radius of the series (\ref{r10}) defined by Hadamard formula as follows:
	\begin{eqnarray*}
		\limsup_{s+t \longrightarrow \infty} \sqrt[s+t]{|r_{st}|R^s_1\,R^t_2}=1.
	\end{eqnarray*}
	For the proof and more details see \cite{TN}. 
	We denote by 
	${\mathcal O}(\mathbb{D}_R)$ the set of holomorphic functions defined
	on $\mathbb{D}_R.$
	We define the  $\mathcal{R}(p,q)$-numbers  \cite{HB}
	\begin{eqnarray}\label{rpqnumbers}
	[n]_{\mathcal{R}(p,q)}:=\mathcal{R}(p^n,q^n),\quad n\in\mathbb{N},
	\end{eqnarray}
	the
	$\mathcal{R}(p,q)$-factorials 
	\begin{eqnarray*}\label{s0}
		[n]!_{\mathcal{R}(p,q)}:=\left \{
		\begin{array}{l}
			1\quad\mbox{for}\quad n=0\\
			\\
			\mathcal{R}(p,q)\cdots\mathcal{R}(p^n,q^n)\quad\mbox{for}\quad n\geq 1,
		\end{array}
		\right .
	\end{eqnarray*}
	and the  $\mathcal{R}(p,q)$-binomial coefficients
	\begin{eqnarray*}\label{bc}
		\bigg[\begin{array}{c} m  \\ n\end{array} \bigg]_{\mathcal{R}(p,q)} := \frac{[m]!_{\mathcal{R}(p,q)}}{[n]!_{\mathcal{R}(p,q)}[m-n]!_{\mathcal{R}(p,q)}},\quad m,n\in\mathbb{N}\cup\{0\},\quad m\geq n.
	\end{eqnarray*}
	
	The  algebra associated with the $\mathcal{R}(p,q)$-deformation is a quantum algebra, denoted $\mathcal{A}_{\mathcal{R}(p,q)},$ generated by the set of operators $\{1, A, A^{\dagger}, N\}$ satisfying the following commutation relations:
	\begin{eqnarray}
	&& \label{algN1}
	\quad A A^\dag= [N+1]_{\mathcal {R}(p,q)},\quad\quad\quad A^\dag  A = [N]_{\mathcal {R}(p,q)}.
	\cr&&\left[N,\; A\right] = - A, \qquad\qquad\quad \left[N,\;A^\dag\right] = A^\dag
	\end{eqnarray}
	with the realization on  ${\mathcal O}(\mathbb{D}_R)$ given by:
	\begin{eqnarray*}\label{algNa}
		A^{\dagger} := z,\qquad A:=\partial_{\mathcal {R}(p,q)}, \qquad N:= z\partial_z,
	\end{eqnarray*} 
	where $\partial_z:=\frac{\partial}{\partial z}$ is the  derivative on $\mathbb{C}.$
	
From these generalized quantum algebras, many works have been carried out. 
	 For instance, the $\mathcal{R}(p,q)$-deformations of
	orthogonal polynomials and univariate discrete distributions of the probability theory (binomial, Euler, P\'olya and inverse P\'olya) were defined and discussed. Moreover, the relevant $\mathcal{R}(p,q)$-factorial moments of a
	random variable and their properties (recurrence relations, mean and variance) were  established \cite{HMD}.  
	The multivariate probability distributions (P\'olya, inverse P\'olya, hypergeometric and negative hypergeometric) were constructed. Furthermore, the corresponding  bivariate probability distributions and   their properties were determined and discussed in \cite{Melong1}. 
    Further, the multinomial coefficients and their recurrence relations as well as the $\mathcal{R}(p,q)$-multinomial probability distribution, the negative $\mathcal{R}(p,q)$-deformed multinomial probability distribution, and the recurrence relations were derived in \cite{Melong2}. In the same vein, 
     the trinomial probability distribution of the first and second kinds, their $\mathcal{R}(p,q)$-factorial moments and  corresponding covariance were investigated in \cite{Melong3}.
	
	Our present work aims at generalizing the multivariate uniform probability distributions and their related properties (mean, variance and covariance) in the framework of $\mathcal{R}(p,q)$-quantum algebras \cite{HB1}.
	
	The paper is organized as follows: In section $2,$ 
	preliminary notions on the multivariate hypergeometric sums from the $\mathcal{R}(p,q)$-quantum algebras are presented. Section $3$ is dedicated  to the 
	construction of the $\mathcal{R}(p,q)$-multivariate uniform probability distributions of the first and second kinds. Their bivariate probability distributions and some properties (mean, variance and covariance) are derived. Particular cases induced by known quantum algebras  are deduced in Section 4. Finally, we end with concluding remarks in section $5.$ 
	\section{Generalized multivariate hypergeometric sums}
	In this section, we investigate two multivariate $\mathcal{R}(p,q)$-hypergeometric sums used in the sequel. The specific numbers  emerging  from the study of $\mathcal{R}(p,q)$-analogues of the Fermi-Dirac and
	Bose-Einstein stochastic models in the sense of statistics are contained in the following results:  
	\begin{proposition}\label{c1}
		Let $k$ and $n$ be positive integers, $p$ and $q$ be  real number, such that $0<q<p\leq 1.$ Then the following relations hold:
		\begin{eqnarray}\label{hs1}
		\sum \tau^{-\sum_{j=1}^{k}jr_j+{n\choose 2}}_1\tau^{\sum_{j=1}^{k}jr_j-{n\choose 2}}_2=\left[\begin{array}{c} k+1  \\ n\end{array} \right]_{\mathcal{R}(p,q)},\quad n\leq k+1, 
		\end{eqnarray}
		where  the summation is over  $r_{j}\in\{0,1\},$ and $j\in\{1,2,\ldots,k\},$ with $\sum_{j=1}^{k}r_j\leq n$ and 
		\begin{eqnarray}\label{hs2}
		\sum \tau^{-\sum_{j=1}^{k}jr_j}_1\tau^{\sum_{j=1}^{k}jr_j}_2=\left[\begin{array}{c} k+n  \\ n\end{array} \right]_{\mathcal{R}(p,q)}. 
		\end{eqnarray}
		Here, the summation is over  $r_{j}\in\{0,1,\ldots,n\},$ and $j\in\{1,2,\ldots,k\},$ with $\sum_{j=1}^{k}r_j\leq n.$
	\end{proposition}
	Two  multivariate $\mathcal{R}(p,q)$-hypergeometric sums are determined in the following 
	theorem.
	\begin{theorem}
		The following results hold:
		\begin{eqnarray}\label{hsa}
		\left[\begin{array}{c} k+1  \\ n\end{array} \right]_{\mathcal{R}(p,q)}=\sum \frac{\tau^{\sum_{j=1}^{\nu}(n-s_j)(m_j-r_j)}_1}{\tau^{\sum_{j=1}^{\nu}(n_j+n-s_j-k-1))r_j}_2}\prod_{j=1}^{\nu}\left[\begin{array}{c} m_j  \\ r_j\end{array} \right]_{\mathcal{R}(p,q)} ,
		\end{eqnarray}
		where $r_j\in\{0,1,\ldots,m_j\}\,,j\in\{0,1,\ldots,\nu\}, \,\sum_{j=1}^{\nu}r_j\leq n $ and $s_j=\sum_{i=1}^{j}r_i,$ and
		\begin{eqnarray}\label{hsb}
		\left[\begin{array}{c} k+n  \\ n\end{array} \right]_{\mathcal{R}(p,q)}
		=\sum \frac{\tau^{\sum_{j=1}^{\nu}(n-s_j)(m_j-1)}_1}{\tau^{\sum_{j=1}^{\nu}(n_j-k-1))r_j}_2}\prod_{j=1}^{\nu}\left[\begin{array}{c} m_j +r_j-1 \\ r_j\end{array} \right]_{\mathcal{R}(p,q)},
		\end{eqnarray}
	with $r_j\in\{0,1,\ldots,n\}\,,j\in\{0,1,\ldots,\nu\}, \, r_1+r_2+\ldots+r_{\nu}\leq n.$
	\end{theorem}
{\it Proof.}  
	From the $\mathcal{R}(p,q)$-Cauchy's formula, the following relation is true:\begin{align*}
	\left[\begin{array}{c} k-n_{j-1}+1  \\ n-s_{j-1}\end{array} \right]_{\mathcal{R}(p,q)}&=\sum_{r_j=0}^{n-s_{j-1}}\tau^{(k-n_j-n+s_j+1))r_j}_2\tau^{(n-s_j)(m_j-r_j)}_1\\&\times\left[\begin{array}{c} m_j  \\ r_j\end{array} \right]_{\mathcal{R}(p,q)}\left[\begin{array}{c} k-n_j+1  \\ n-s_j\end{array} \right]_{\mathcal{R}(p,q)},
	\end{align*}
	where $j\in\{1,2,\ldots,\nu\}.$ Starting with the first expression, $j=1,$
	\begin{align*}
	\left[\begin{array}{c} k+1  \\ n\end{array} \right]_{\mathcal{R}(p,q)}&=\sum_{r_1=0}^{n-s}\tau^{(k-n_1-n+s_1+1))r_1}_2\tau^{(n-s_1)(r_1-m_1)}_1\\&\times\left[\begin{array}{c} m_1  \\ r_1\end{array} \right]_{\mathcal{R}(p,q)}\left[\begin{array}{c} k-n_1+1  \\ n-s_1\end{array} \right]_{\mathcal{R}(p,q)},
	\end{align*}and replacing the second factor of the general term of the sum by
\begin{align*}
\left[\begin{array}{c} k-n_1+1  \\ n-s_1\end{array} \right]_{\mathcal{R}(p,q)}&=\sum_{r_2=0}^{n-r_1}\tau^{(k-n_2-n+s_2+1))r_2}_2\tau^{(n-s_2)(r_2-m_2)}_1\\&\times\left[\begin{array}{c} m_2  \\ r_2\end{array} \right]_{\mathcal{R}(p,q)}\left[\begin{array}{c} k-n_2+1  \\ n-s_2\end{array} \right]_{\mathcal{R}(p,q)},
\end{align*}
and, continuing in this manner, at the last step replacing the second factor of the general
term of the sum by
\begin{align*}
	\left[\begin{array}{c} k-n_{\nu-1}+1  \\ n-s_{\nu-1}\end{array} \right]_{\mathcal{R}(p,q)}&=\sum_{r_{\nu}=0}^{n-s_{\nu-1}}\tau^{(k-n_{\nu}-n+s_{\nu}+1))r_{\nu}}_2\tau^{(n-s_{\nu})(r_{\nu}-m_{\nu})}_1\\&\times\left[\begin{array}{c} m_{\nu}  \\ r_{\nu}\end{array} \right]_{\mathcal{R}(p,q)}\left[\begin{array}{c} k-n_{\nu}+1  \\ n-s_{\nu}\end{array} \right]_{\mathcal{R}(p,q)}\\&=\sum_{r_{\nu}=0}^{n-s_{\nu-1}}\frac{\tau^{(k-n_{\nu}-n+s_{\nu}+1))r_{\nu}}_2}{\tau^{(s_{\nu}-n)(r_{\nu}-m_{\nu})}_1}\left[\begin{array}{c} m_{\nu}  \\ r_{\nu}\end{array} \right]_{\mathcal{R}(p,q)},
\end{align*}
the relation \eqref{hsa} is deduced. By analogy, from the $\mathcal{R}(p,q)$-Cauchy's formula,  the following formula is also true:
\begin{align*}
\left[\begin{array}{c} k-n_{j-1}+n-s_{j-1}  \\ n-s_{j-1}\end{array} \right]_{\mathcal{R}(p,q)}&=\sum_{r_j=0}^{n-s_{j-1}}\frac{\tau^{(n-s_j)(m_j-1)}_1}{\tau^{(n_j-k-1))r_j}_2}\left[\begin{array}{c} m_j+r_j-1  \\ r_j\end{array} \right]_{\mathcal{R}(p,q)}\\&\times\left[\begin{array}{c} k-n_j+n-s_j  \\ n-s_j\end{array} \right]_{\mathcal{R}(p,q)},
\end{align*}
where $j\in\{1,2,\ldots,\nu\}.$ Starting with the first expression, $j=1,$
\begin{eqnarray*}
	\left[\begin{array}{c} k+n  \\ n\end{array} \right]_{\mathcal{R}(p,q)}&=&\sum_{r_1=0}^{n}\tau^{(n-s_1)(m_1-1)}_1\tau^{(k-n_1+1))r_1}_2\\&\times&\left[\begin{array}{c} m_1+r_1-1  \\ r_1\end{array} \right]_{\mathcal{R}(p,q)}\left[\begin{array}{c} k-n_1+n-s_1  \\ n-s_1\end{array} \right]_{\mathcal{R}(p,q)},
\end{eqnarray*}and replacing the second factor of the general term of the sum by:
\begin{align*}
	\left[\begin{array}{c} k-n_1+n-s_1  \\ n-s_1\end{array} \right]_{\mathcal{R}(p,q)}&=\sum_{r_2=0}^{n-r_1}\frac{\tau^{(n-s_2)(m_2-1)}_1}{\tau^{(n_2-k-1))r_2}_2}\left[\begin{array}{c} m_2+r_2-1  \\ r_2\end{array} \right]_{\mathcal{R}(p,q)}\\&\times\left[\begin{array}{c} k-n_2+n-s_2  \\ n-s_2\end{array} \right]_{\mathcal{R}(p,q)},
\end{align*}
and, continuing in this manner, at the last step replacing the second factor of the general
term of the sum by
\begin{align*}
	\left[\begin{array}{c} k-n_{\nu-1}+n-s_{\nu-1}  \\ n-s_{\nu-1}\end{array} \right]_{\mathcal{R}(p,q)}&=\sum_{r_{\nu}=0}^{n-s_{\nu-1}}\frac{\tau^{(n-s_{\nu})(m_{\nu}-1)}_1}{\tau^{(n_{\nu}-k-1))r_{\nu}}_2}\left[\begin{array}{c} m_{\nu}+r_{\nu}-1  \\ r_{\nu}\end{array} \right]_{\mathcal{R}(p,q)}\\&\times\left[\begin{array}{c} k-n_{\nu}+n-s_{\nu}  \\ n-s_{\nu}\end{array} \right]_{\mathcal{R}(p,q)}\\
&=\sum_{r_{\nu}=0}^{n-s_{\nu-1}}\frac{\tau^{(n-s_{\nu})(m_{\nu}-1)}_1}{\tau^{(n_{\nu}-k-1))r_{\nu}}_2}\left[\begin{array}{c} m_{\nu} +r_{\nu}-1 \\ r_{\nu}\end{array} \right]_{\mathcal{R}(p,q)},
\end{align*}
expression \eqref{hsb} is deduced and the proof is achieved. $\cqfd$
\begin{remark}
	Relevant particular cases of multivariate hypergeometric sums related to quantum algebras in the literature  are deduced as:
	\begin{enumerate}
		\item[(a)]
	Two  multivariate $(p,q)$-hypergeometric sums are deduced by taking $\mathcal{R}(x,y)=\frac{x-y}{p-q},$ implies $\tau_1=p$ and $\tau_2=q:$
	\begin{eqnarray*}
	\sum p^{-\sum_{j=1}^{k}jr_j+{n\choose 2}}\,q^{\sum_{j=1}^{k}jr_j-{n\choose 2}}=\left[\begin{array}{c} k+1  \\ n\end{array} \right]_{p,q},\quad n\leq k+1, 
	\end{eqnarray*}
	where  the summation is over  $r_{j}\in\{0,1\},$ and $j\in\{1,2,\ldots,k\},$ with $\sum_{j=1}^{k}r_j\leq n$ and 
	\begin{eqnarray*}
	\sum p^{-\sum_{j=1}^{k}jr_j}\,q^{\sum_{j=1}^{k}jr_j}=\left[\begin{array}{c} k+n  \\ n\end{array} \right]_{p,q}, 
	\end{eqnarray*}
	  the summation being performed over  $r_{j}\in\{0,1,\ldots,n\},$ and $j\in\{1,2,\ldots,k\},$ with $\sum_{j=1}^{k}r_j\leq n.$ Moreover, we have 
		\begin{equation*}
		\left[\begin{array}{c} k+1  \\ n\end{array} \right]_{p,q}=\sum p^{\sum_{j=1}^{\nu}(n-s_j)(m_j-r_j)}q^{\sum_{j=1}^{\nu}(k-n_j-n+s_j+1))r_j}\prod_{j=1}^{\nu}\left[\begin{array}{c} m_j  \\ r_j\end{array} \right]_{p,q} ,
		\end{equation*}
		where $r_j\in\{0,1,\ldots,m_j\}\,,j\in\{0,1,\ldots,\nu\}, \, r_1+r_2+\ldots+r_{\nu}\leq n$ and $s_j=\sum_{i=1}^{j}r_i,$ and
		\begin{equation*}
		\left[\begin{array}{c} k+n  \\ n\end{array} \right]_{p,q}
		=\sum p^{\sum_{j=1}^{\nu}(n-s_j)(m_j-1)}q^{\sum_{j=1}^{\nu}(k-n_j+1))r_j}\prod_{j=1}^{\nu}\left[\begin{array}{c} m_j +r_j-1 \\ r_j\end{array} \right]_{p,q},
		\end{equation*}
		with $r_j\in\{0,1,\ldots,n\}\,,j\in\{0,1,\ldots,\nu\}, \, r_1+r_2+\ldots+r_{\nu}\leq n.$
		\item[(b)]Two  multivariate generalized $q$- Quesne hypergeometric sums are derived by taking $\mathcal{R}(x,y)=\frac{xy-1}{(q-p^{-1})y},$ implies $\tau_1=p$ and $\tau_2=q^{-1}.$
	\end{enumerate}
\end{remark}
\section{$\mathcal{R}(p,q)$-Fermi-Dirac stochastic model}
In this section, we construct the multivariate uniform probability distribution of the first kind induced from the generalized quantum algebra\cite{HB1}. Moreover, the bivariate case and their properties ($\mathcal{R}(p,q)$-mean, $\mathcal{R}(p,q)$-variance and $\mathcal{R}(p,q)$-covariance) are determined and presented. Particular cases of multivariate uniform probability distributions of the first kind are deduced.

A random distribution (placement) of balls into distinguishable urns (cells) is a simple
and very useful stochastic model. Among its most striking and useful applications, the
Bose-Einstein and Fermi-Dirac stochastic models (statistics) deserve special attention.
A random $\mathcal{R}(p,q)$-distribution (placement) of a ball into $r$ distinguishable urns (cells) $\{c_1, c_2, \ldots , c_r \}$ can be introduced in the following way. Suppose that $r$ numbered balls $\{1, 2, \ldots , r \},$ representing the $r$ urns are forced to pass through a random mechanism, 
one after the other, in the order $\big(1, 2, \ldots , r \big)$ or in the reverse order $\big(r, r-1, \ldots , 2,1 \big).$ Moreover, assume that each passing ball can or cannot be caught by the mechanism,
with probabilities $p = 1-q$ and $q,$ respectively. In the case all $r$ balls pass through
the mechanism and no ball is caught, the ball passing the procedure is repeated, with the
same order. Then, the number on the first caught ball determines the urn (cell) in which
the ball is placed. Clearly, the probability that a ball is placed in the $j^{th}$ order urn
is given by:
\begin{eqnarray*}
P_j=\sum_{k=0}^{\infty}(\tau_1-\tau_2)\tau^{(j-1)+kr}_2=\frac{\tau^{j-1}_2}{[r]_{\mathcal{R}(p,q)}},\quad j\in\{1,2,\ldots,r\},
\end{eqnarray*}
or
\begin{eqnarray*}
P_j=\sum_{k=0}^{\infty}(\tau_1-\tau_2)\tau^{(r-j)+kr}_2=\frac{\tau^{r-j}_2}{[r]_{\mathcal{R}(p,q)}},\quad j\in\{1,2,\ldots,r\},
\end{eqnarray*}
according to whether the ball passing order is $(1, 2,\ldots, r )$
or $(r , r-1, \ldots , 2, 1).$
These probabilities, by using the expression $\frac{(\tau_1\tau_2)^{j-1} 
}{[r ]_{\mathcal{R}(p,q)}} =
\frac{(\tau_1\tau_2)^{j-r} 
}{[r ]_{\mathcal{R}(p^{-1}, q^{-1})}},$ can be written in a single formula as follows:
\begin{eqnarray*}
	P_j=\frac{\tau^{r-j}_2}{[r]_{\mathcal{R}(p,q)}},\quad j\in\{1,2,\ldots,r\}.
\end{eqnarray*}
Note that this is the probability function of a
discrete $\mathcal{R}(p,q)$-uniform distribution of the set $\{1, 2, \ldots, r \}.$ 

Moreover, suppose  that $n$ indistinguishable balls are randomly $\mathcal{R}(p,q)$-distributed, one after the
other, into $r=k+1,$ distinguishable urns (cells) $\{c_1,c_2,\ldots,c_{k+1}\},$ each with capacity limited to one ball, with $n\leq k+1.$

 We denote by $X_j$  the number of balls placed in urn $c_j,$ for $j\in\{1,2,\ldots,k+1\}.$ Note that $X_{k+1}=n-X_1-X_2-\ldots X_k.$ The distribution of the random vector
$\underline{X}=(X_1, X_2, \ldots, X_k )$ is called {\it multivariate discrete $\mathcal{R}(p,q)$-uniform distribution of the first
	kind,} with parameters $n,$ $p$ and $q.$ Its probability function is derived in the following
theorem.
\begin{theorem}
	The probability function of the multivariate discrete $\mathcal{R}(p,q)$-uniform distribution of the first
	kind, with parameters $n,$ $p$ and $q,$ is given by:
	\begin{equation}\label{rpqmufk}
	P\big(X_1=x_1,X_2=x_2,\ldots,X_k=x_k, \big)=\frac{\Phi(p,q)}{\left[\begin{array}{c} k+1  \\ n\end{array} \right]_{\mathcal{R}(p,q)}},
	\end{equation}
	where $\Phi(p,q)=\tau^{-\sum_{j=1}^{k}(k-j+1)x_j+{n\choose 2}+kn}_1\tau^{\sum_{j=1}^{k}(k-j+1)x_j-{n\choose 2}}_2,\, \sum_{j=1}^{k}x_j\leq n,\,x_j\in\{0,1\}$ and $j\in\{1,2,\ldots,k\}.$
\end{theorem}
{\it Proof.} 
	By using the corollary \eqref{c1}. $\cqfd$

The multivariate discrete $\mathcal{R}(p,q)$-uniform distribution of the first kind can be obtained as
the conditional distribution of $k$ independent $\mathcal{R}(p,q)$-Bernoulli distributions of the first kind,
given their sum with another $\mathcal{R}(p,q)$-Bernoulli distribution of the first kind independent of them.

We consider a sequence of independent Bernoulli trials and suppose that the
probability of success at the $i^{th}$ trial is given by:
\begin{equation}
P_{i}=\frac{\theta\tau^{i-1}_2}{\tau^{i-1}_1+ \theta\tau^{i-1}_2},\quad i\in\mathbb{N}.
\end{equation}
Note that, taking $\mathcal{R}(1,x)=\frac{1-x}{1-q},$ implies $\tau_1=1$ and $\tau_2=q,$ we recover the reults given in\cite{CA2022}:
\begin{equation*}
P_{i}=\frac{\theta\,q^{i-1}}{1+ \theta\,q^{i-1}},\quad i\in\mathbb{N}.
\end{equation*}
For $j\in\{1,2,\ldots,k+1\},$ we denote by $X_j$  the number of successes at the $j^{th}$ trial.
\begin{theorem}
	The
	conditional probability function of the random vector $\underline{X}=\big(X_1,X_2,\ldots,X_k\big)$ given that $\sum_{i=1}^{k+1}X_i=n,$ is the $\mathcal{R}(p,q)$-multivariate uniform distribution of the first kind with mass function \eqref{rpqmufk}.
\end{theorem}
{\it Proof.}  
	The random variables $X_j, j\in\{1, 2, \ldots , k+1\},$ are independent, with probability
	function  given by: 
	\begin{equation*}
	P\big(X_j=x_j\big)=\frac{\theta^{x_j}\tau^{(j-1)x_j}_2\tau^{(j-1)(1-x_j)}_1}{\tau^{j-1}_1+ \theta\tau^{j-1}_2},\quad x_j\in\{0,1\}.
	\end{equation*}
	By analogy, the probability function of the sum $Y_{k+1}=\sum_{i=1}^{k+1}X_i,$ which
	is the number of successes in $k+1$ trials, is given by:
	\begin{equation*}
	P\big(Y_k+1=n\big)=\left[\begin{array}{c}k+1 \\n\end{array} \right]_{\mathcal{R}(p,q)}\frac{\theta^n\tau^{k+1-n\choose 2}_1\tau^{n\choose 2}_2}{\big(1\oplus \theta\big)^{k+1}_{\mathcal{R}(p,q)}},\quad n\in\{0,1,\ldots,k+1\}.
	\end{equation*}
	Then, the joint conditional probability function of the  vector $\big(X_1,X_2,\ldots,X_k\big),$ given that $Y_k+1=n,$ is 
	\begin{align*}
	P\big(X_1=x_1, X_2=&x_2,\ldots, X_k=x_k|Y_{k+1}=n\big)\\&=\frac{P(X_1=x_1)\ldots P(X_k=x_k)P(X_{k+1}=n-y_k)}{P\big(Y_{k+1}=n\big)},
	\end{align*}
where $y_k=\sum_{j=1}^{k}x_j.$ By using these expressions, we obtain:
\begin{equation}\label{ri}
P\big(X_1=x_1, X_2=x_2,\ldots, X_k=x_k|Y_{k+1}=n\big)=\frac{\tau^{\beta(n,k,\underline{x})}_1\tau^{\alpha(n,k,\underline{x})}_2}{\left[\begin{array}{c}k+1 \\n\end{array} \right]_{\mathcal{R}(p,q)}},
\end{equation}
where 
\begin{align*}
\alpha(n,k,\underline{x})&=\sum_{j=1}^{k}(j-1)x_j-\sum_{j=1}^{k}kx_j+nk-{n\choose 2}\\&=-\sum_{j=1}^{k}(k-j+1)x_j+{n\choose 2}+n(k-n+1)
\end{align*}
and 
\begin{align*}
\beta(n,k,\underline{x})&=\sum_{j=1}^{k}(j-1)(1-x_j)+\sum_{j=1}^{k}kx_j+k(1-n)-{k+1-n\choose 2}\\&=\sum_{j=1}^{k}\big((k-j+1)x_j+(j-1)\big)+k(1-n)-{k+1-n\choose 2}\\&\sum_{j=1}^{k}(k-j+1)x_j +{n\choose 2} +n(1-n).
\end{align*}
By using the relation
\begin{equation}
\big(\tau_1\tau_2\big)^{-n(k-n+1)}\left[\begin{array}{c}k+1 \\n\end{array} \right]_{\mathcal{R}(p,q)}=\left[\begin{array}{c}k+1 \\n\end{array} \right]_{\mathcal{R}(p^{-1},q^{-1})},
\end{equation}
the equation \eqref{ri} is reduced to  
\begin{equation}
P\big(X_1=x_1,\ldots, X_k=x_k|Y_{k+1}=n\big)=\Phi(n,k)\frac{1}{\left[\begin{array}{c}k+1 \\n\end{array} \right]_{\mathcal{R}(p^{-1},q^{-1})}},
\end{equation}
where $\Phi(n,k)=\tau^{\sum_{j=1}^{k}(k-j+1)x_j+{n\choose 2}-kn}_1\tau^{-\sum_{j=1}^{k}(k-j+1)x_j+{n\choose 2}}_2.$ Thus, the relation \eqref{rpqmufk} follows by replacing $\mathcal{R}(p,q)$ by $\mathcal{R}(p^{-1},q^{-1}).$ 

 Some marginal and conditional distributions of the multivariate $\mathcal{R}(p,q)$-uniform distribution
of the first kind are deduced in the following theorem.
\begin{theorem}
	Suppose that the random vector $\underline{X}=\big(X_1, X_2,\ldots,X_k\big)$ satisfies a multivariate
	discrete $\mathcal{R}(p,q)$-uniform distribution of the first kind. Then:
	\begin{enumerate}
		\item[(i)] The probability function of the marginal distribution of $\big(X_1, X_2,\ldots,X_r\big)$ for $1\leq r< k,$ is given by:
		\begin{eqnarray}\label{i1}
		P\big(X_1,X_2,\ldots,X_r\big)= \Phi_1(p,q)\frac{\left[\begin{array}{c}k-r+1 \\n-y_r\end{array} \right]_{\mathcal{R}(p,q)}}{\left[\begin{array}{c}k+1 \\n\end{array} \right]_{\mathcal{R}(p,q)}},
		\end{eqnarray}
		where $\Phi_1(p,q)=\tau^{-\sum_{j=1}^{r}(k-j-n+y_r+1)x_j+{y_r\choose 2}+kn}_1\tau^{\sum_{j=1}^{r}(k-j-n+y_r+1)x_j-{y_r\choose 2}}_2,$  $x_j\in\{0,1\}, j\in\{1,2,\ldots,r\},$ with $\sum_{j=1}^{r}x_j\leq n, y_r=\sum_{j=1}^{r}x_j.$
		\item[(ii)]The conditional  probability distribution of the vector $\big(X_{r+1}, X_{r+2},\ldots,X_{r+m}\big),$ given that $\big(X_1, X_2,\ldots,X_r\big)=\big(x_1, x_2,\ldots,x_r\big)$ for $1\leq r<m\leq k,$ is given by:
		\begin{align}\label{ii1}
		P\big(X_{r+1}=x_{r+1},\ldots,X_m=x_m|X_1&=x_1,\ldots,X_r=x_r\big)= \Phi_2(p,q)\nonumber\\&\times\frac{\left[\begin{array}{c}k-m+1 \\n-y_m\end{array} \right]_{\mathcal{R}(p,q)}}{\left[\begin{array}{c}k-r+1 \\n-y_r\end{array} \right]_{\mathcal{R}(p,q)}},
		\end{align}
		where \begin{align*}\Phi_2(p,q)&=\tau^{-\sum_{j=r+1}^{m}(k-j-n+y_r+1)x_j+{y_m-y_r\choose 2}+kn}_1\\&\times\tau^{\sum_{j=r+1}^{m}(k-j-n+y_m+1)x_j-{y_m-y_r\choose 2}}_2,\end{align*} $x_j\in\{0,1\}, j\in\{r+1,r+2,\ldots,m\},$ with $\sum_{j=r+1}^{m}x_j\leq n-y_r, y_j=\sum_{i=1}^{j}x_i.$
	\end{enumerate}
\end{theorem} 
{\it Proof.}
	\begin{enumerate}
		\item[(i)]Summing the probability function of the multivariate discrete $\mathcal{R}(p,q)$-uniform
		distribution of the first kind, for $x_j\in\{0,1\},\,j\in\{r+1,r+2,\ldots,k\},\, \sum_{j=r+1}^{k}x_j\leq n-y_r$ and using the relation
		\begin{equation*}
		{n\choose 2}={n-y_r\choose 2}+{y_r\choose 2}+y_r(n-y_r),
		\end{equation*}
		we get, for the marginal probability function of $(X_1, X_2, \ldots, X_r ),$ the expression
		\begin{align*}
		P\big(X_1=x_1,\ldots,X_r=x_r\big)&=\tau^{-\sum_{j=1}^{r}(k-n-j+y_r+1)x_j+{y_r\choose 2}+kn}_1\\&\times\tau^{\sum_{j=1}^{r}(k-n-j+y_r+1)x_j-{y_r\choose 2}}_2\sum \frac{\phi(j,k-r)}{\left[\begin{array}{c} k+1  \\ n\end{array} \right]_{\mathcal{R}(p,q)}},
		\end{align*}
		where $\phi(j,k-r)=\frac{\tau^{\sum_{j=1}^{k-r}(k-r-j+1)x_{r+j}-{n-y_r\choose 2}}_2}{\tau^{\sum_{j=1}^{k-r}(k-r-j+1)x_{r+j}+{n-y_r\choose 2}}_1},$  $x_j\in\{0,1\},\,j\in\{1,2,\ldots,k-r\},\, \sum_{j=r+1}^{k}x_j\leq n-y_r.$ Since, the multiple sum, using \eqref{hs1}, equals
		\begin{eqnarray*}
		\sum\tau^{-\sum_{j=1}^{k-r}(k-r-j+1)x_{r+j}+{n-y_r\choose 2}}_2 \tau^{\sum_{j=1}^{k-r}(k-r-j+1)x_{r+j}-{n-y_r\choose 2}}_2=\left[\begin{array}{c} k-r+1  \\ n-y_r\end{array} \right]_{\mathcal{R}(p,q)},
		\end{eqnarray*}
	with $x_j\in\{0,1\},\,j\in\{1,2,\ldots,k-r\},\, \sum_{j=r+1}^{k}x_j\leq n-y_r,$ the last expression of probability function reduces to \eqref{i1}.
		\item[(ii)]The conditional probability of $(X_{r+1}, X_{r+2}, \ldots , X_m),$ given that
		$(X_1, X_2, \ldots , X_r ) = (x_1, x_2, \ldots, x_r ),$ is given by
		\begin{align*}
		P\big(X_{r+1}=x_{r+1},\ldots, X_m=x_m|X_1&=x_1,\ldots X_{r}=x_{r}\big)\\&=\frac{P\big(X_1=x_1, X_2=x_2\ldots X_m=x_m\big)}{P\big(X_1=x_1, X_2=x_2\ldots X_{r}=x_{r}\big)}.
		\end{align*}
		Then, using the result of part \eqref{i1}, together with the following relation
		\begin{equation*}
		{y_m-y_r\choose 2}={y_m\choose 2}-{y_r\choose 2}-y_r(y_m-y_r),
		\end{equation*}
		we deduce that 
		\begin{align*}
		P\big(X_{r+1}=x_{r+1},\ldots, X_m=x_m|X_1=x_1,\ldots &,X_{r}=x_{r}\big)=\Phi(r+1,m)\\&\times\frac{\left[\begin{array}{c} k-m+1  \\ n-y_m\end{array} \right]_{\mathcal{R}(p,q)}}{\left[\begin{array}{c} k-r+1  \\ n-y_r\end{array} \right]_{\mathcal{R}(p,q)}},
		\end{align*}
		where \begin{align*}\Phi(r+1,m)&=\tau^{-\sum_{j=r+1}^{m}(k-j-n+y_m+1)x_j+{y_m-y_r\choose 2}+kn}_1\\&\times\tau^{\sum_{j=r+1}^{m}(k-j-n+y_m+1)x_j-{y_m-y_r\choose 2}}_2\end{align*} $x_j\in\{0,1\},\,j\in\{r+1,r+2,\ldots,m\},\, \sum_{j=r+1}^{m}x_j\leq n-y_r$ and $y_j=\sum_{i=1}^{j}x_i.$
	\end{enumerate}

Now, we consider the random variables 
\begin{eqnarray}
Y_j=\sum_{i=s_{j-1}+1}^{s_j} X_i=\sum_{i=1}^{m_j}X_{s_{j-1}}+i,\quad j\in\{1,2,\ldots,r\},
\end{eqnarray}
where $m_i,i\in\{1,2,\ldots,r\}$ are positive integers and $s_j=\sum_{i=1}^{j}m_j, j\in\{1,2,\ldots,r\},$ with $s_r=k,$ and $s_0=0.$

We assume that the random vector $(X_1, X_2,\ldots, X_k )$ satisfies a multivariate
discrete $\mathcal{R}(p,q)$-uniform distribution of the first kind. Then, 
the probabilistic behaviour of groups of successive urns
(energy levels) is determined in the following theorem.
\begin{theorem}
	\begin{enumerate}
		\item[(i)] 
		The probability function of
		the distribution of the random vector $(Y_1, Y_2, \ldots, Y_r )$ is given by:
		\begin{align}\label{i}
			P\big(Y_1=y_1,Y_2=y_2,\ldots,Y_r&=y_r\big)=\tau^{\sum_{j=1}^{r}(n-z_j-s_j)(m_j-y_j)}_1\nonumber\\&\times\tau^{\sum_{j=1}^{r}(k-s_j-n+z_j+1)y_j}_2\frac{\prod_{j=1}^{r}\left[\begin{array}{c} m_j  \\ y_j\end{array} \right]_{\mathcal{R}(p,q)}}{\left[\begin{array}{c} k+1  \\ n\end{array} \right]_{\mathcal{R}(p,q)}},
		\end{align}
		where $y_j\in\{0,1,\ldots,m_j\},\, z_j=\sum_{i=1}^{j}y_i,\,j\in\{1,2,\ldots,r\},\,\sum_{j=1}^{r}y_j\leq n.$
		\item[(ii)]The probability function of the marginal distribution of the  vector $(Y_1, Y_2,\ldots , Y_{\nu} ),$ for $1\leq \nu < r,$ is given by:
		\begin{align}\label{ii}
			P\big(Y_1=y_1,\ldots,Y_{\nu}=y_{\nu}\big)&=\frac{\tau^{\sum_{j=1}^{\nu}(n-z_j-s_j)(m_j-y_j)}_1}{\tau^{\sum_{j=1}^{\nu}(s_j-n+z_j-k-1)y_j}_2}\prod_{j=1}^{r}\left[\begin{array}{c} m_j  \\ y_j\end{array} \right]_{\mathcal{R}(p,q)}\nonumber\\&\times\frac{\left[\begin{array}{c} k-s_{\nu}+1  \\ n-z_{\nu}\end{array} \right]_{\mathcal{R}(p,q)} }{\left[\begin{array}{c} k+1  \\ n\end{array} \right]_{\mathcal{R}(p,q)}},
		\end{align}
		where $\sum_{j=1}^{r}y_j\leq n,\,y_j\in\{0,1,\ldots,m_j\},\, z_j=\sum_{i=1}^{j}y_i,\,j\in\{1,2,\ldots,r\}.$
		\item[(iii)]The probability of  the conditional distribution of the vector $(Y_{\nu+1}, Y_{\nu+2},\ldots , Y_{\nu} ),$ given
		that $(Y_1, Y_2,\ldots , Y_{\nu} ) = (y_1, y_2, \ldots, y_{\nu} ),$ for $1\leq \nu <k\leq r,$ is given by
		\begin{align}\label{iii}
			P\big(Y_{\nu+1}=y_{\nu+1},\ldots,&Y_{k}=y_{k}|Y_1=y_1,\ldots,Y_{\nu}=y_{\nu}\big)=\Phi(\nu+1,k)\nonumber\\&\times\prod_{j=\nu+1}^{k}\left[\begin{array}{c} m_j  \\ y_j\end{array} \right]_{\mathcal{R}(p,q)}\frac{\left[\begin{array}{c} k-s_{k}+1  \\ n-z_{k}\end{array} \right]_{\mathcal{R}(p,q)} }{\left[\begin{array}{c} k-s_{\nu}+1  \\ n-z_{\nu}\end{array} \right]_{\mathcal{R}(p,q)}},
		\end{align}
	where $\Phi(\nu+1,k)=\tau^{\sum_{j=\nu+1}^{k}(n-z_j-s_j)(m_j-y_j)}_1\tau^{\sum_{j=1}^{\nu}(k-s_j+n-z_j+1)y_j}_2$ $\sum_{j=\nu+1}^{k}y_j\leq n-z_{\nu},\,y_j\in\{0,1,\ldots,n-z_{\nu}\},\, z_j=\sum_{i=1}^{j}y_i,\,j\in\{\nu+1,\ldots,k\}.$
	\end{enumerate}
\end{theorem}
{\it Proof.} 
\begin{enumerate}
	\item[(i)] The probability function of the random vector $(Y_1, Y_2, \ldots, Y_r )$ is deduced
	by using  the probability function
	\begin{eqnarray*}
	P\big(X_1=x_1,\ldots,X_k=x_k\big)=\frac{\tau^{-\sum_{j=1}^{k}(k-j+1)x_j+{n\choose 2}+kn}_1\tau^{\sum_{j=1}^{k}(k-j+1)x_j-{n\choose 2}}_2}{\left[\begin{array}{c} k+1  \\ n\end{array} \right]_{\mathcal{R}(p,q)}},
	\end{eqnarray*}
	 inserting into it the $r$ variables $(y_1, y_2, \ldots, y_r )$ and summing the new
	expression over all the remaining $k-r$ old variables. Note that,
	\begin{eqnarray*}
		\sum_{j=1}^{r}y_j=\sum_{j=1}^{r}\sum_{i=s_{j-1}+1}^{s_j}x_i=\sum_{i=1}^{k}x_i.
	\end{eqnarray*}
	The sum in the exponent of $\tau_1$ and $\tau_2$ can be expressed as follows:
	\begin{eqnarray*}
		\sum_{j=1}^{r}(k-j+1)x_j=\sum_{j=1}^{r}\sum_{i=s_{j-1}+1}^{s_j}(k-i+1)x_i=\sum_{j=1}^{r}\sum_{i=s_{j-1}+1}^{s_{j-1}+m_j}(k-i+1)x_i.
	\end{eqnarray*}
	Moreover, replacing in the last inner sum  the variable $i$ by $s_{j-1}+i$ and inserting
	into the resulting expression the variables $(y_1, y_2, \ldots, y_r ),$ we get
	\begin{align*}
		\sum_{j=1}^{k}(k-j+1)x_j&=\sum_{j=1}^{r}\sum_{i=1}^{m_j}(k-s_{j-1}-i+1)x_{s_{j-1}+i}\\&=\sum_{j=1}^{r}(k-s_j+1)\sum_{i=1}^{m_j}x_{s_{j-1}+i}+\sum_{j=1}^{r}\sum_{i=1}^{m_j-1}(m_j-1)x_{s_{j-1}+i}\\&=\sum_{j=1}^{r}(k-s_j+1)y_j+\sum_{j=1}^{r}\sum_{i=1}^{m_j-1}(m_j-1)x_{s_{j-1}+i}.
	\end{align*}
	Thus, the probability function of the random vector $(Y_1, Y_2, \ldots, Y_r )$ is given by
	\begin{align*}
		P\big(Y_1=y_1,\ldots,Y_r=y_r\big)&=\frac{\Phi(j,r)}{\left[\begin{array}{c} k+1  \\ n\end{array} \right]_{\mathcal{R}(p,q)}}\sum\tau^{\sum_{j=1}^{r}\sum_{i=1}^{m_j-1}(m_j-1)x_{s_{j-1}+i}}_2\\&\times \tau^{-\sum_{j=1}^{r}\sum_{i=1}^{m_j-1}(m_j-1)x_{s_{j-1}+i}}_1\\&=\frac{\Phi(j,r)}{\left[\begin{array}{c} k+1  \\ n\end{array} \right]_{\mathcal{R}(p,q)}}\prod_{j=1}^{r}\sum\tau^{-\sum_{i=1}^{m_j-1}(m_j-1)x_{s_{j-1}+i}}_1\\&\times\tau^{\sum_{i=1}^{m_j-1}(m_j-1)x_{s_{j-1}+i}}_2 ,
	\end{align*}
	where $\Phi(j,r)=\tau^{\sum_{j=1}^{r}(k-s_j+1)y_j-{n\choose 2}+kn}_1\tau^{\sum_{j=1}^{r}(k-s_j+1)y_j-{n\choose 2}}_2,$ the summation, in the last sum, is extended over all $x_{s_{j-1}+i}\in\{0,1\}$ for $i\in\{0,1,\ldots,m_{j}-1\},$ with $\sum_{i=1}^{m_j-1}x_{s_{j-1}+i}\leq y_j.$ In addition, to these values, the summation
	in the first sum is extended to all $j\in\{1,2,\ldots r\}.$ Moreover, by using \eqref{hs1}
	\begin{eqnarray*}
		\sum\tau^{-\sum_{i=1}^{m_j-1}(m_j-i)x_{s_{j-1}+i}+{y_j\choose 2}}_1\tau^{\sum_{i=1}^{m_j-1}(m_j-i)x_{s_{j-1}+i}-{y_j\choose 2}}_2=\left[\begin{array}{c} m_j  \\ y_j\end{array} \right]_{\mathcal{R}(p,q)},
	\end{eqnarray*}
	where the summation is over $x_{s_{j-1}+i}\in\{0,1\},\,i\in\{0,1,\ldots,m_{j}-1\},$ with $x_{s_{j-1}+1}+x_{s_{j-1}+2}+\ldots+ x_{s_{j-1}+m_j-1}\leq y_j.$ Moreover, summing the relations
	\begin{eqnarray*}
		{n-z_j \choose 2}={n-z_{j-1} \choose 2}-{y_j \choose 2}-y_j(n-z_j),\quad j\in\{1,2,\ldots,r\},
	\end{eqnarray*}
	where $z_j=\sum_{i=1}^{j}y_i,\, j\in\{1,2,\ldots,r\},\,z_0=0,$ and since
	\begin{align*}
		{n-z_r\choose 2}&={n-y_1-y_2-\ldots-y_r\choose 2}\\&={n-x_1-x_2-\ldots-x_k\choose 2}\\&={x_{k+1}\choose 2}=0,\quad x_{k+1}\in\{0,1\},
	\end{align*}
	we have 
	\begin{equation*}
		{n\choose 2}-\sum_{j=1}^{r}{y_j\choose 2}=\sum_{j=1}^{r}(n-z_j)y_j.
	\end{equation*}
	Putting into the last expression of the probability $P(Y_1 = y_1, Y_2 = y_2, \ldots , Y_r = y_r )$ these two above
	expressions, and after computation, we obtain \eqref{i}.
	\item[(ii)]Summing the probability function of the random vector $(Y_1, Y_2, \ldots, Y_r ),$ for $y_j\in\{0,1,\ldots,m_j\},\,j\in\{\nu+1,\nu+2,\ldots,r\},$ with $\sum_{j=\nu+1}^{r}y_j\leq n-z_r,$
	\begin{align*}
		P\big(Y_1=y_1,Y_2=y_2,\ldots,Y_{\nu}=&y_{\nu}\big)=\frac{\tau^{\sum_{j=\nu+1}^{\nu}(n-z_j-s_j)(m_j-y_j)}_1}{\tau^{\sum_{j=1}^{\nu}(s_j+n-z_j-k-1)y_j}_2}\\&\times\frac{\prod_{j=1}^{\nu}\left[\begin{array}{c} m_j  \\ y_j\end{array} \right]_{\mathcal{R}(p,q)}}{\left[\begin{array}{c} k+1  \\ n\end{array} \right]_{\mathcal{R}(p,q)}}\\&\times\frac{\tau^{\sum_{j=\nu+1}^{\nu}(n-z_j-s_j)(m_j-y_j)}_1}{\tau^{\sum_{j=\nu+1}^{r}(s_j+n-z_j-k-1)y_j}_2}\prod_{j=\nu+1}^{r}\left[\begin{array}{c} m_j  \\ y_j\end{array} \right]_{\mathcal{R}(p,q)},
	\end{align*}
	where $y_{\nu+j}\in\{0,1,\ldots,m_j\},\,j\in\{1,2,\ldots,r-\nu\},\,y_{\nu+1}+\ldots+y_r\leq n-z_{\nu}$ and using \eqref{hsa},
	\begin{eqnarray*}
		\left[\begin{array}{c} k-s_{\nu}+1  \\ n-z_{\nu}\end{array} \right]_{\mathcal{R}(p,q)}&=\sum \frac{\tau^{\sum_{j=\nu+1}^{\nu}(n-z_j-s_j)(m_j-y_j)}_1}{\tau^{\sum_{j=\nu+1}^{r}(s_j+n-z_j-k-1))y_j}_2}\prod_{j=\nu+1}^{r}\left[\begin{array}{c} m_j  \\ y_j\end{array} \right]_{\mathcal{R}(p,q)} ,
	\end{eqnarray*}
	with $y_{\nu+j}\in\{0,1,\ldots,m_j\},\,j\in\{1,2,\ldots,r-\nu\},\,y_{\nu+1}+\ldots+y_r\leq n-z_{\nu},$ the probability function \eqref{ii} is obtained.
	\item[(iii)]The conditional probability function of $(Y_{\nu+1}, Y_{\nu+2},\ldots , Y_k ),$ given that
	$$(Y_1, Y_2, \ldots, Y_{\nu} ) = (y_1, y_2,\ldots, y_{\nu} ),$$ is given by:
	\begin{align*}
		P\big(Y_{\nu+1}=y_{\nu+1},\ldots, Y_k=y_k|Y_1&=y_1,\ldots Y_{\nu}=y_{\nu}\big)\\&=\frac{P\big(Y_1=y_1, Y_2=y_2\ldots Y_k=y_k\big)}{P\big(Y_1=y_1, Y_2=y_2\ldots Y_{\nu}=y_{\nu}\big)}.
	\end{align*}
	Then, by using the realtions \eqref{i} and \eqref{ii}, the formula is readily deduced.$\cqfd$
\end{enumerate}
\subsection{$\mathcal{R}(p,q)$-bivariate uniform distribution of the first kind}
Now, we study  the moments of a multivariate $\mathcal{R}(p,q)$-uniform distribution
of the first kind. For the computation of the $\mathcal{R}(p,q)$-power moments, and particularly
 the $\mathcal{R}(p,q)$-mean, $\mathcal{R}(p,q)$-variance, and $\mathcal{R}(p,q)$-covariance, the attention can be focussed
to the marginal distribution of the random vector $\underline{X}=(X_1, X_2).$ Putting $r=2$ in the relation \eqref{i1}, we deduce  
	the probability function of the $\mathcal{R}(p,q)$-bivariate uniform distribution of the first kind as: 
	\begin{equation}\label{rpqbufk}
	P\big(X_1=x_1, X_2=x_2\big)=\Phi(p,q)\frac{\left[\begin{array}{c} k-1  \\ n-x_1-x_2\end{array} \right]_{\mathcal{R}(p,q)}}{\left[\begin{array}{c} k+1  \\ n\end{array} \right]_{\mathcal{R}(p,q)}}, 
	\end{equation}
	where 
	$\Phi_1(p,q)=\tau^{-(k-n+x_1)x_1-(k-n+x_1+x_2-1)x_2+kn}_1\tau^{(k-n+x_1)x_1+(k-n+x_1+x_2-1)x_2}_2,$
 $x_1\in\{0,1\}, x_2\in\{0,1\},  $ with $x_1+x_2\leq n.$

 The $\mathcal{R}(p^{-1},q^{-1})$-power moments of the
 random vector $\big(X_1,X_2\big)$ are given in the following theorem. 
 \begin{theorem}
 	We assume that the probability function of the random vector $\big(X_1,X_2\big)$ is given by \eqref{rpqbufk}. Then, the following results hold:
 	\begin{equation}\label{rpqbufka}
 	E\big([X]^{i_1}_{\mathcal{R}(p^{-1},q^{-1})}\big)=\tau^{kn}_1\frac{[n
 		]_{\mathcal{R}(p^{-1},q^{-1})}}{[k+1]_{\mathcal{R}(p^{-1},q^{-1})}}, \quad i_1\in\mathbb{N}
 	\end{equation}
 	and
 	\begin{align}\label{rpqbufkb}
 	V\big([X_1]_{\mathcal{R}(p^{-1},q^{-1})}\big)
 	=\frac{\tau^{kn}_1[n
 		]_{\mathcal{R}(p^{-1},q^{-1})}}{[k+1]_{\mathcal{R}(p^{-1},q^{-1})}}\bigg(1-\frac{\tau^{kn}_1[n
 		]_{\mathcal{R}(p^{-1},q^{-1})}}{[k+1]_{\mathcal{R}(p^{-1},q^{-1})}}\bigg).
 	\end{align}
 	Furthermore,\begin{equation*}
 	E\big(\tau^{-X_1}_2[X_2]^{i_2}_{\mathcal{R}(p^{-1},q^{-1})}\big)=\frac{\tau^{-1}_2\,[n]_{\mathcal{R}(p^{-1},q^{-1})}}{\tau^{n(1-k)}_1[k+1]_{\mathcal{R}(p^{-1},q^{-1})}},\quad i_2\in\mathbb{N}\cup\{0\}
 	\end{equation*}
 	and
 	\begin{align}
 	Cov\big([X_1]_{\mathcal{R}(p^{-1},q^{-1})},\tau^{-X_1}_2[X_2]_{\mathcal{R}(p^{-1},q^{-1})}\big)&=\frac{\tau^{-1}_2[n]_{\mathcal{R}(p^{-1},q^{-1})}}{\tau^{n(1-k)}_1[k+1]^2_{\mathcal{R}(p^{-1},q^{-1})}}\nonumber\\&\times \frac{\Delta(p,q)}{[k]_{\mathcal{R}(p^{-1},q^{-1})}},
 	\end{align}
 	where $$\Delta(p,q)=\tau^{n}_1[n-1]_{\mathcal{R}(p^{-1},q^{-1})}[k+1]_{\mathcal{R}(p^{-1},q^{-1})}-[n]_{\mathcal{R}(p^{-1},q^{-1})}[k]_{\mathcal{R}(p^{-1},q^{-1})}.$$
 \end{theorem} 
{\it Proof.}
	The marginal probability function of the random variable $X_1$ is given by:
	\begin{eqnarray*}
	P\big(X_1=x_1\big)=\tau^{-(k-n+x_1)x_1+kn}_1\tau^{(k-n+x_1)x_1}_2\frac{\left[\begin{array}{c} k-1  \\ n-x_1\end{array} \right]_{\mathcal{R}(p,q)}}{\left[\begin{array}{c} k+1  \\ n\end{array} \right]_{\mathcal{R}(p,q)}},\quad x_1\in\{0,1\}.
	\end{eqnarray*}
	We  focus on  $\mathcal{R}(p^{-1},q^{-1})$-factorial moments. Then, using the relation
	\begin{equation*}
	\big(\tau_1\tau_2\big)^{-n(k-n)}\left[\begin{array}{c}k \\n\end{array} \right]_{\mathcal{R}(p,q)}=\left[\begin{array}{c}k \\n\end{array} \right]_{\mathcal{R}(p^{-1},q^{-1})},
	\end{equation*}
	the above relation can be written as:
	\begin{eqnarray}
	P\big(X_1=x_1\big)=\tau^{n(x_1-1)+kn}_1\tau^{n(1-x_1)}_2\frac{\left[\begin{array}{c} k-1  \\ n-x_1\end{array} \right]_{\mathcal{R}(p^{-1},q^{-1})}}{\left[\begin{array}{c} k+1  \\ n\end{array} \right]_{\mathcal{R}(p^{-1},q^{-1})}},\quad x_1\in\{0,1\}.
	\end{eqnarray}
	Naturally, the $\mathcal{R}(p^{-1},q^{-1})$-power moments of the random variable $X_1$ is obtained as follows:
	\begin{small}
	\begin{align*}
	E\big([X]^{i_1}_{\mathcal{R}(p^{-1},q^{-1})}\big)&=\sum_{x_1=0}^{1}\tau^{n(x_1-1)+kn}_1\tau^{n(1-x_1)}_2 [x]^{i_1}_{\mathcal{R}(p^{-1},q^{-1})}\frac{\left[\begin{array}{c} k  \\ n-x_1\end{array} \right]_{\mathcal{R}(p^{-1},q^{-1})}}{\left[\begin{array}{c} k+1  \\ n\end{array} \right]_{\mathcal{R}(p^{-1},q^{-1})}}\\&=\tau^{kn}_1\frac{\left[\begin{array}{c} k  \\ n-1\end{array} \right]_{\mathcal{R}(p^{-1},q^{-1})}}{\left[\begin{array}{c} k+1  \\ n\end{array} \right]_{\mathcal{R}(p^{-1},q^{-1})}}.
	\end{align*}
\end{small}
	By using the  relation
	\begin{equation*}
	\left[\begin{array}{c} k+1  \\ n\end{array} \right]_{\mathcal{R}(p^{-1},q^{-1})}=\frac{[k+1]_{\mathcal{R}(p^{-1},q^{-1})}}{[n]_{\mathcal{R}(p^{-1},q^{-1})}}\left[\begin{array}{c} k  \\ n-1\end{array} \right]_{\mathcal{R}(p^{-1},q^{-1})},
	\end{equation*}
the equation \eqref{rpqbufka} holds. Moreover, the $\mathcal{R}(p^{-1},q^{-1})$-variance of the random variable $X_1$ is derived as:
\begin{align*}
V\big([X_1]_{\mathcal{R}(p^{-1},q^{-1})}\big)&=E\big([X_1]^{2}_{\mathcal{R}(p^{-1},q^{-1})}\big)- \big[E\big([X_1]_{\mathcal{R}(p^{-1},q^{-1})}\big)\big]^2\\
&=\frac{[n
	]_{\mathcal{R}(p^{-1},q^{-1})}}{[k+1]_{\mathcal{R}(p^{-1},q^{-1})}}\bigg(1-\frac{[n
	]_{\mathcal{R}(p^{-1},q^{-1})}}{[k+1]_{\mathcal{R}(p^{-1},q^{-1})}}\bigg).
\end{align*}
 Moreover, the expected value of $\tau^{-X_1}_2[X_2]^{i_2}_{\mathcal{R}(p^{-1},q^{-1})}$ can be obtained by using the relation
\begin{eqnarray*}
E\big(\tau^{-X_1}_2[X_2]^{i_2}_{\mathcal{R}(p^{-1},q^{-1})}\big)=E\big[E\big(\tau^{-X_1}_2[X_2]^{i_2}_{\mathcal{R}(p^{-1},q^{-1})}\big)/X_1\big].
\end{eqnarray*}
The conditional probability function of $X_2,$ given that $X_1 = x_1,$ is provided by 
	\begin{eqnarray*}
P\big(X_2=x_2|X_1=x_1\big)=\tau^{(n-x_1)(x_2-1)+kn}_1\tau^{(n-x_1)(1-x_2)}_2\frac{\left[\begin{array}{c} k-1  \\ n-x_1-x_2\end{array} \right]_{\mathcal{R}(p^{-1},q^{-1})}}{\left[\begin{array}{c} k  \\ n-x_1\end{array} \right]_{\mathcal{R}(p^{-1},q^{-1})}},
\end{eqnarray*}
where $x_2\in\{0,1\}$ is the  probability function of $X_1,$ with the parameters $k$ and $n$
replaced by $k-1$ and $n-x_1,$ respectively. Then, 
\begin{eqnarray*}
E\big([X_2]^{i_2}_{\mathcal{R}(p^{-1},q^{-1})}|X_1=x_1\big)=\frac{\tau^{kn}_1[n-x_1
	]_{\mathcal{R}(p^{-1},q^{-1})}}{[k]_{\mathcal{R}(p^{-1},q^{-1})}}.
\end{eqnarray*}
Besides, the expected value of $\tau^{-X_1}_2[n-X_1]_{\mathcal{R}(p^{-1},q^{-1})}$ is given by:
\begin{align*}
E\big(\tau^{-X_1}_2[n-X_1]^{i_2}_{\mathcal{R}(p^{-1},q^{-1})}\big)&=\sum_{x_1=0}^{1}\tau^{n(x_1-1)+kn}_1\tau_2^{n(1-x_1)-x_1}\\&\times	[n-x_1]_{\mathcal{R}(p^{-1},q^{-1})}\frac{\left[\begin{array}{c} k  \\ n-x_1\end{array} \right]_{\mathcal{R}(p^{-1},q^{-1})}}{\left[\begin{array}{c} k+1  \\ n\end{array} \right]_{\mathcal{R}(p^{-1},q^{-1})}}\\
&=[k]_{\mathcal{R}(p^{-1},q^{-1})}\frac{\bigg(\tau^{n(k-1)}_1\tau_2^{n}\left[\begin{array}{c} k-1  \\ n-1\end{array} \right]_{\mathcal{R}(p^{-1},q^{-1})}}{\left[\begin{array}{c} k+1  \\ n\end{array} \right]_{\mathcal{R}(p^{-1},q^{-1})}}\\&+\frac{\tau^{kn}_1\tau^{-1}_2\left[\begin{array}{c} k-1  \\ n-2\end{array} \right]_{\mathcal{R}(p^{-1},q^{-1})}\bigg)}{\left[\begin{array}{c} k+1  \\ n\end{array} \right]_{\mathcal{R}(p^{-1},q^{-1})}}.
\end{align*}
Using the triangular recurrence relation of the $\mathcal{R}(p,q)$-binomial coefficients, we obtain:
\begin{align*}
E\big(\tau^{X_1}_2[n-X_1]_{\mathcal{R}(p^{-1},q^{-1})}\big)
&=\frac{\tau^{-1}_2[k]_{\mathcal{R}(p^{-1},q^{-1})}\tau_2\left[\begin{array}{c} k  \\ n-1\end{array} \right]_{\mathcal{R}(p^{-1},q^{-1})}}{\tau^{n(1-k)}_1\left[\begin{array}{c} k+1  \\ n\end{array} \right]_{\mathcal{R}(p^{-1},q^{-1})}}\\&=\frac{\tau^{-1}_2\,[n]_{\mathcal{R}(p^{-1},q^{-1})}[k]_{\mathcal{R}(p^{-1},q^{-1})}}{\tau^{n(1-k)}_1[k+1]_{\mathcal{R}(p^{-1},q^{-1})}}.
\end{align*}
Thus,
\begin{align*}
E\big(\tau^{-X_1}_2[X_2]^{i_2}_{\mathcal{R}(p^{-1},q^{-1})}\big)&=E\big[E\big(\tau^{-X_1}_2[X_2]^{i_2}_{\mathcal{R}(p^{-1},q^{-1})}|X_1\big)\big]\\&=\frac{1}{[k]_{\mathcal{R}(p^{-1},q^{-1})}}\,E\big(\tau^{-X_1}_2[n-X_1]_{\mathcal{R}(p^{-1},q^{-1})}\big)\\&=\frac{\tau^{-1}_2\,[n]_{\mathcal{R}(p^{-1},q^{-1})}}{\tau^{n(1-k)}_1[k+1]_{\mathcal{R}(p^{-1},q^{-1})}}.
\end{align*}
By analogy, the expected value of the function
$\tau^{-X_1}_2[X_1]^{i_1}_{\mathcal{R}(p^{-1},q^{-1})}[X_2]^{i_2}_{\mathcal{R}(p^{-1},q^{-1})}$ can be computed by using the relation
\begin{eqnarray*}
E\big(\tau^{-X_1}_2[X_1]^{i_1}_{\mathcal{R}(p^{-1},q^{-1})}[X_2]^{i_2}_{\mathcal{R}(p^{-1},q^{-1})}\big)=E\big[E\big(\tau^{-X_1}_2[X_1]^{i_1}_{\mathcal{R}(p^{-1},q^{-1})}[X_2]^{i_2}_{\mathcal{R}(p^{-1},q^{-1})}|X_1\big)\big].
\end{eqnarray*}
Clearly, \begin{align*}
E\big(\tau^{-X_1}_2[X_1]^{i_1}_{\mathcal{R}(p^{-1},q^{-1})}[n-X_1&]^{i_2}_{\mathcal{R}(p^{-1},q^{-1})}\big)=\sum_{x_1=0}^{1}\tau^{(x_1-1)n+kn}_1\tau_2^{n(1-x_1)-x_1}[x_1]^{i_1}_{\mathcal{R}(p^{-1},q^{-1})}\\&\times	[n-x_1]_{\mathcal{R}(p^{-1},q^{-1})}\frac{\left[\begin{array}{c} k  \\ n-x_1\end{array} \right]_{\mathcal{R}(p,q)}}{\left[\begin{array}{c} k+1  \\ n\end{array} \right]_{\mathcal{R}(p,q)}}\\
&=[k]_{\mathcal{R}(p^{-1},q^{-1})}\sum_{x_1=0}^{1}\tau^{(x_1-1)n+kn}_1\tau_2^{n(1-x_1)-x_1}\\&\times	[x_1]^{i_1}_{\mathcal{R}(p^{-1},q^{-1})}\frac{\left[\begin{array}{c} k  \\ n-x_1-1\end{array} \right]_{\mathcal{R}(p^{-1},q^{-1})}}{\left[\begin{array}{c} k+1  \\ n\end{array} \right]_{\mathcal{R}(p^{-1},q^{-1})}}\\
&=[k]_{\mathcal{R}(p^{-1},q^{-1})}\tau^{kn}_1\tau^{-1}_2\frac{\left[\begin{array}{c} k-1  \\ n-2\end{array} \right]_{\mathcal{R}(p^{-1},q^{-1})}}{\left[\begin{array}{c} k+1  \\ n\end{array} \right]_{\mathcal{R}(p^{-1},q^{-1})}}.
\end{align*}
and
\begin{equation*}
E\big(\tau^{-X_1}_2[X_1]^{i_1}_{\mathcal{R}(p^{-1},q^{-1})}[n-X_1]^{i_2}_{\mathcal{R}(p^{-1},q^{-1})}\big)
=\frac{\tau^{kn}_1\tau^{-1}_2\,[n]_{2,\mathcal{R}(p^{-1},q^{-1})}}{[k+1]_{\mathcal{R}(p^{-1},q^{-1})}}.
\end{equation*}
Then,
\begin{small}
\begin{align*}
	E\bigg(\frac{\tau^{X_1}_2}{[X_1]^{i_1}_{\mathcal{R}(p^{-1},q^{-1})}}[X_2]^{i_2}_{\mathcal{R}(p^{-1},q^{-1})}\bigg)&=E\bigg[E\bigg(\frac{\tau^{X_1}_2}{[X_1]^{i_1}_{\mathcal{R}(p^{-1},q^{-1})}}[X_2]^{i_2}_{\mathcal{R}(p^{-1},q^{-1})}|X_1\bigg)\bigg]\\&=\frac{E\big(\tau^{X_1}_2[X_1]^{i_1}_{\mathcal{R}(p^{-1},q^{-1})}[n-X_1]_{\mathcal{R}(p^{-1},q^{-1})}\big)}{[k]_{\mathcal{R}(p^{-1},q^{-1})}}\\&=\frac{\tau^{kn}_1\tau^{-1}_2\,[n]_{2,\mathcal{R}(p^{-1},q^{-1})}}{[k+1]_{2,\mathcal{R}(p^{-1},q^{-1})}}.
\end{align*}
\end{small}
The covariance of the random variable $[X_1]_{\mathcal{R}(p^{-1},q^{-1})}$ and $\tau^{-X_1}_2[X_2]_{\mathcal{R}(p^{-1},q^{-1})}$ is given by:
\begin{align*}
Cov\big([X_1]_{\mathcal{R}(p^{-1},q^{-1})},\tau^{-X_1}_2[X_2]_{\mathcal{R}(p^{-1},q^{-1})}\big)&=\frac{\tau^{kn}_1\tau^{-1}_2\,[n]_{\mathcal{R}(p^{-1},q^{-1})}}{[k+1]_{\mathcal{R}(p^{-1},q^{-1})}[k]_{\mathcal{R}(p^{-1},q^{-1})}}\\&-\frac{\tau^{-1}_2\,[n]^2_{\mathcal{R}(p^{-1},q^{-1})}[n-1]_{\mathcal{R}(p^{-1},q^{-1})}}{\tau^{n(1-k)}_1[k+1]^2_{\mathcal{R}(p^{-1},q^{-1})}}.
\end{align*}
$\cqfd$
\subsection{Particular cases of multivariate uniform distribution of first kind} We derive from the general formalism the multivariate uniform probability distribution of the first kind and related properties generated by the quantum algebras existing in the literature.
	\begin{enumerate}
		\item[(a)]The multivariate uniform probability distribution of the first kind and properties from the {\bf Arick-Coon-Kuryskin algebra}\cite{AC} can be obtained by  putting $\mathcal{R}(x)=\frac{x-x^{-1}}{q-q^{-1}}.$
		\item[(b)]
	The multivariate uniform probability distribution and properties corresponding to the {\bf Jagannathan  and  Srinivasa algebra} \cite{JS} are deduced by taking $\mathcal{R}(x,y)=\frac{x-y}{p-q}$: The probability function of the multivariate discrete $\mathcal{R}(p,q)$-uniform distribution of the first
	kind, with parameters $n,$ $p$ and $q,$ is given by:
	\begin{equation*}
	P\big(X_1=x_1,X_2=x_2,\ldots,X_k=x_k, \big)=\frac{\Phi(p,q)}{\left[\begin{array}{c} k+1  \\ n\end{array} \right]_{p,q}},
	\end{equation*}
	where $\Phi(p,q)=p^{-\sum_{j=1}^{k}(k-j+1)x_j+{n\choose 2}+kn}\,q^{\sum_{j=1}^{k}(k-j+1)x_j-{n\choose 2}},\, x_j\in\{0,1\},\,j\in\{1,2,\ldots,k\}$ and $\sum_{j=1}^{k}x_j\leq n.$ Besides, we suppose that the random vector $\underline{X}=\big(X_1, X_2,\ldots,X_k\big)$ satisfies a multivariate
	discrete $(p,q)$-uniform distribution of the first kind. Then:
	\begin{enumerate}
		\item[(i)] The probability function of the marginal distribution of the random vector $\big(X_1, X_2,\ldots,X_r\big)$ for $1\leq r< k,$ is given by:
		\begin{eqnarray*}
		P\big(X_1,X_2,\ldots,X_r\big)= \frac{q^{\sum_{j=1}^{r}(k-j-n+y_r+1)x_j-{y_r\choose 2}}\left[\begin{array}{c}k-r+1 \\n-y_r\end{array} \right]_{p,q}}{p^{\sum_{j=1}^{r}(k-j-n+y_r+1)x_j-{y_r\choose 2}-kn}\left[\begin{array}{c}k+1 \\n\end{array} \right]_{p,q}},
		\end{eqnarray*}
		where   $x_j\in\{0,1\}, j\in\{1,2,\ldots,r\},$ with $\sum_{j=1}^{r}x_j\leq n, y_r=\sum_{j=1}^{r}x_j.$
		\item[(ii)]The conditional  probability distribution of the  vector $\big(X_{r+1}, \ldots,X_{r+m}\big),$ provided that $\big(X_1, X_2,\ldots,X_r\big)=\big(x_1, x_2,\ldots,x_r\big)$ for $1\leq r<m\leq k,$ yields:
		\begin{align*}
			P\big(X_{r+1}=x_{r+1},\ldots,&X_m=x_m|X_1=x_1,\ldots,X_r=x_r\big)=\nonumber\\& \frac{q^{\sum_{j=r+1}^{m}(k-j-n+y_m+1)x_j-{y_m-y_r\choose 2}}\left[\begin{array}{c}k-m+1 \\n-y_m\end{array} \right]_{p,q}}{p^{\sum_{j=r+1}^{m}(k-j-n+y_r+1)x_j-{y_m-y_r\choose 2}-kn}\left[\begin{array}{c}k-r+1 \\n-y_r\end{array} \right]_{p,q}},
		\end{align*}
		where $x_j\in\{0,1\}, j\in\{r+1,r+2,\ldots,m\},$ with $\sum_{j=r+1}^{m}x_j\leq n-y_r, y_j=\sum_{i=1}^{j}x_i.$
	\end{enumerate}
	We consider the random variables 
	\begin{eqnarray*}
	Y_j=\sum_{i=s_{j-1}+1}^{s_j} X_i=\sum_{i=1}^{m_j}X_{s_{j-1}}+i,\quad j\in\{1,2,\ldots,r\},
	\end{eqnarray*}
	where $m_i,i\in\{1,2,\ldots,r\}$ are positive integers and $s_j=\sum_{i=1}^{j}m_j, j\in\{1,2,\ldots,r\},$ with $s_r=k,$ and $s_0=0.$
	
	We assume that the random vector $(X_1, X_2,\ldots, X_k )$ obeys a multivariate
	discrete $(p,q)$-uniform distribution of the first kind. Then:
		\begin{enumerate}
			\item[(i)] 
			The probability function of
			the distribution of the random vector $(Y_1, Y_2, \ldots, Y_r )$ is expressed as:
			\begin{equation*}
			P\big(Y_1=y_1,Y_2=y_2,\ldots,Y_r=y_r\big)=\frac{p^{\sum_{j=1}^{r}(n-z_j-s_j)(m_j-y_j)}\prod_{j=1}^{r}\left[\begin{array}{c} m_j  \\ y_j\end{array} \right]_{p,q}}{q^{-\sum_{j=1}^{r}(k-s_j-n+z_j+1)y_j}\left[\begin{array}{c} k+1  \\ n\end{array} \right]_{p,q}},
			\end{equation*}
			where   $y_j\in\{0,1,\ldots,m_j\},\, z_j=\sum_{i=1}^{j}y_i,\,j\in\{1,2,\ldots,r\},\,\sum_{j=1}^{r}y_j\leq n.$
			\item[(ii)]The probability function of the marginal distribution of the random vector $(Y_1, Y_2,\ldots , Y_{\nu} ),$ for $1\leq \nu < r,$ is furnished by:
			\begin{align*}
			P\big(Y_1=y_1,Y_2=y_2,\ldots,Y_{\nu}=y_{\nu}\big)&=\frac{p^{\sum_{j=1}^{\nu}(n-z_j-s_j)(m_j-y_j)}}{q^{-\sum_{j=1}^{\nu}(k-s_j+n-z_j+1)y_j}}\nonumber\\&\times\prod_{j=1}^{r}\left[\begin{array}{c} m_j  \\ y_j\end{array} \right]_{p,q}\frac{\left[\begin{array}{c} k-s_{\nu}+1  \\ n-z_{\nu}\end{array} \right]_{p,q} }{\left[\begin{array}{c} k+1  \\ n\end{array} \right]_{p,q}},
			\end{align*}
			where $y_j\in\{0,1,\ldots,m_j\},\, z_j=\sum_{i=1}^{j}y_i,\,j\in\{1,2,\ldots,r\},$ and $\sum_{j=1}^{r}y_j\leq n.$
			\item[(iii)]The probability function of  the conditional distribution of the random vector $(Y_{\nu+1}, Y_{\nu+2},\ldots , Y_{\nu} ),$ given
			that $(Y_1, Y_2,\ldots , Y_{\nu} ) = (y_1, y_2, \ldots, y_{\nu} ),$ for $1\leq \nu <k\leq r,$ is afforded by
			\begin{align*}
			P\big(Y_{\nu+1}=y_{\nu+1},\ldots,Y_{k}=&y_{k}|Y_1=y_1,Y_2=y_2,\ldots,Y_{\nu}=y_{\nu}\big)=\nonumber\\&\times \frac{p^{\sum_{j=\nu+1}^{k}(n-z_j-s_j)(m_j-y_j)}}{q^{-\sum_{j=\nu+1}^{k}(k-s_j+n-z_j+1)y_j}}\nonumber\\&\times\prod_{j=\nu+1}^{k}\left[\begin{array}{c} m_j  \\ y_j\end{array} \right]_{p,q}\frac{\left[\begin{array}{c} k-s_{k}+1  \\ n-z_{k}\end{array} \right]_{p,q} }{\left[\begin{array}{c} k-s_{\nu}+1  \\ n-z_{\nu}\end{array} \right]_{p,q}},
			\end{align*}
		\end{enumerate}

	 Moreover, the properties of the $(p,q)$-bivariate uniform probability distribution such as the $(p,q)$-mean, $(p,q)$-variance and $(p,q)$-covariance are expressed by:
	\begin{equation*}
	E\big([X]^{i_1}_{p^{-1},q^{-1}}\big)=\frac{p^{kn}\,[n
		]_{p^{-1},q^{-1}}}{[k+1]_{p^{-1},q^{-1}}}, \quad i_1\in\mathbb{N}
	\end{equation*}
	and
	\begin{equation*}
	V\big([X_1]_{p^{-1},q^{-1}}\big)
	=\frac{p^{kn}\,[n
		]_{p^{-1},q^{-1}}}{[k+1]_{p^{-1},q^{-1}}}\bigg(1-\frac{p^{kn}\,[n
		]_{p^{-1},q^{-1}}}{[k+1]_{p^{-1},q^{-1}}}\bigg).
	\end{equation*}
	Furthermore,\begin{equation*}
	E\big(q^{-X_1}[X_2]^{i_2}_{p^{-1},q^{-1}}\big)=\frac{q\,[n]_{p^{-1},q^{-1}}}{p^{n(1-k)}[k+1]_{p^{-1},q^{-1}}},\quad i_2\in\mathbb{N}\cup\{0\}
	\end{equation*}
	and
	\begin{eqnarray*}
	Cov\big([X_1]_{p^{-1},q^{-1}},q^{-X_1}\,[X_2]_{p^{-1},q^{-1}}\big)=\frac{q^{-1}\,[n]_{p^{-1},q^{-1}}\Delta(p,q)}{p^{n(1-k)}[k+1]^2_{p^{-1},q^{-1}}[k]_{p^{-1},q^{-1}}},
	\end{eqnarray*}
	where $$\Delta(p,q)=p^{n}\,[n-1]_{p^{-1},q^{-1}}[k+1]_{p^{-1},q^{-1}}-[n]_{p^{-1},q^{-1}}[k]_{p^{-1},q^{-1}}.$$
	\item[(c)] Setting $\mathcal{R}(x,y)=\frac{1-xy}{(p^{-1}-q)x},$ we deduce the multivariate uniform probability distribution of the first king and properties induced by the {\bf Chakrabarty and Jagannathan algebra}\cite{CJ}.
	\item[(d)]
	The multivariate uniform probability distribution of the first king and properties associated to the {\bf generalized $q$-Quesne  algebra}\cite{HB1} are derived by taking $\mathcal{R}(x,y)=\frac{xy-1}{(q^{-1}-p)y}$:
\end{enumerate}
\section{$\mathcal{R}(p,q)$-Bose-Einstein stochastic model}
This section addresses the multivariate discrete uniform probability distribution of the second kind  from the $\mathcal{R}(p,q)$-deformed quantum algebra \cite{HB}. The $\mathcal{R}(p,q)$-bivariate discrete uniform is derived, and the properties ($\mathcal{R}(p,q)$-mean, $\mathcal{R}(p,q)$-varaince and $\mathcal{R}(p,q)$-covariance) are computed. Besides,  particular cases related to known quantum algebras  are deduced. 

Assume now that $n$ indistinguishable balls are randomly $\mathcal{R}(p,q)$-distributed, one after the
other, into $r=k+1$ distinguishable urns (cells) $\{c_1,c_2,\ldots,c_k\},$ with unlimited
capacity. Let $X_j$ be the number of balls placed in urn $c_j,$ for $j\in\{1,2,\ldots,k+1\}.$ Note that $X_{k+1}=n-X_1-X_2-\ldots X_k.$ The distribution of the random vector
$(X_1, X_2, \ldots, X_k )$ may be  called {\it multivariate discrete $\mathcal{R}(p,q)$-uniform distribution of the second
	kind,} with parameters $n,$ $p$ and $q.$ Its probability function is derived in the following
theorem.

\begin{theorem}
	The probability function of the multivariate discrete $\mathcal{R}(p,q)$-uniform distribution of the second
	kind, with parameters $n,$ $p$ and $q,$ is given by:
\begin{eqnarray}\label{rpqmusk}
P\big(X_1=x_1, X_2=x_2,\ldots, X_k=x_k\big)=\Phi(n,k)\frac{1}{\left[\begin{array}{c}k+n \\n\end{array} \right]_{\mathcal{R}(p^{-1},q^{-1})}},
\end{eqnarray}
where $\Phi(n,k)=\tau^{-\sum_{j=1}^{k}(k-j+1)x_j+2kn+ {k+1\choose 2}}_1\tau^{\sum_{j=1}^{k}(k-j+1)x_j}_2.$

\end{theorem} 	

We consider a sequence of independent Bernoulli trials and suppose that the
conditional probability of success at a trial, given that $j-1$ successes occur in the
previous trials, is given as follows:
\begin{eqnarray*}
P_j=1-\theta\tau^{j-1}_2\tau^{1-j}_1,\quad j\in\mathbb{N}\quad\mbox{and}\quad 0<\theta<1.
\end{eqnarray*}

We denote by $W_j$  the number of failures after the $(j- 1)^{th}$ success and
until the occurrence of the $j^{th}$ success, for $j \in\{1,2,\ldots,k+1\}.$

The multivariate discrete $\mathcal{R}(p,q)$-uniform distribution of the second kind can be presented
as the conditional distribution of $k$ independent $\mathcal{R}(p,q)$-geometric distributions of the second
kind, given their sum with another $\mathcal{R}(p,q)$-geometric distribution of the second kind
independent of them. The result is contained in the next theorem.
\begin{theorem}
	The conditional probability distribution of the  vector $\big(W_1,W_2,\ldots,W_k\big),$ given that $W_1+W_2+\ldots+W_k=n,$ is the multivariate discrete
	$\mathcal{R}(p,q)$-uniform distribution of the second kind with probability function \eqref{rpqmusk}.
\end{theorem}
{\it Proof.}
	Naturally, the random variable $W_j, j\in\{1,2,\ldots,k+1\},$ are independent with probability function
	\begin{eqnarray*}
	P(W_j=wj)=\tau^{(1-j)(w_j+1)}_1\big(\theta\tau^{j-1}_2\big)^{w_j}\big(\tau^{j-1}_1-\theta\tau^{j-1}_2\big),\quad w_j\in\mathbb{N},
	\end{eqnarray*}
where $j\in\{1,2,\ldots,k+1\}.$ Moreover, the probability function of the sum $U_{k+1} = W_1 + W_2 +\ldots+ W_{k+1},$ which is
the number of failures until the occurrence of the $(k + 1)^{th}$ success, is determined by:
\begin{eqnarray*}
P\big(U_{k+1} =n\big)=\left[\begin{array}{c} k+n  \\ n\end{array} \right]_{\mathcal{R}(p,q)}\theta^{n}\prod_{i=1}^{k+1}\big(\tau^{i-1}_1-\theta\tau^{i-1}_2\big),\quad n\in\mathbb{N}.
\end{eqnarray*}
Thus, the joint conditional probability distribution of the vector $(W_1,W_2,\ldots ,W_k ),$
given that $U_{k+1} = n,$ can be expressed as:
\begin{align*}
P\big(W_1=w_1, W_2&=w_2,\ldots, W_k=w_k|U_{k+1}=n\big)\nonumber\\&=\frac{P(W_1=w_1)\ldots P(W_k=w_k)P(W_{k+1}=n-u_k)}{P\big(U_{k+1}=n\big)},
\end{align*}
where $u_k=\sum_{j=1}^{k}w_j.$  Using these expressions, we obtain:
\begin{equation}\label{ri1}
P\big(W_1=w_1, W_2=w_2,\ldots, W_k=w_k|U_{k+1}=n\big)=\frac{\tau^{\eta(n,k,\underline{w})}_1\tau^{\gamma(n,k,\underline{w})}_2}{\left[\begin{array}{c}k+n \\n\end{array} \right]_{\mathcal{R}(p,q)}},
\end{equation}
where 
\begin{equation*}
\gamma(n,k,\underline{w})=\sum_{j=1}^{k}(j-1)w_j-\sum_{j=1}^{k}kw_j+nk=-\sum_{j=1}^{k}(k-j+1)w_j+nk
\end{equation*}
and 
\begin{align*}
\eta(n,k,\underline{w})&=\sum_{j=1}^{k}(1-j)(w_j+1)+\sum_{j=1}^{k}kw_j-k(n+1)\\&=\sum_{j=1}^{k}\big((k-j+1)w_j-(j-1)\big)-k(n+1)\\&=\sum_{j=1}^{k}(k-j+1)w_j-kn- {k+1\choose 2}.
\end{align*}
From the relation
\begin{equation*}
\big(\tau_1\tau_2\big)^{-nk}\left[\begin{array}{c}k+n \\n\end{array} \right]_{\mathcal{R}(p,q)}=\left[\begin{array}{c}k+n \\n\end{array} \right]_{\mathcal{R}(p^{-1},q^{-1})},
\end{equation*}
the expression \eqref{ri1} is reduced to  
\begin{equation*}
P\big(W_1=w_1, W_2=w_2,\ldots, W_k=w_k|U_{k+1}=n\big)=\Phi(n,k)\frac{1}{\left[\begin{array}{c}k+n \\n\end{array} \right]_{\mathcal{R}(p^{-1},q^{-1})}},
\end{equation*}
where $\Phi(n,k)=\tau^{\sum_{j=1}^{k}(k-j+1)w_j-2kn- {k+1\choose 2}}_1\tau^{-\sum_{j=1}^{k}(k-j+1)w_j}_2.$ Thus, the relation \eqref{rpqmusk} follows by replacing $\mathcal{R}(p,q)$ by $\mathcal{R}(p^{-1},q^{-1}).$ 

Another marginal and conditional distributions of the multivariate discrete  $\mathcal{R}(p,q)$-uniform distribution
of the second kind are determined in the  theorem bellow.
\begin{theorem}
	Suppose that the random vector $\big(X_1, X_2,\ldots,X_k\big)$ satisfies a multivariate
	discrete $\mathcal{R}(p,q)$-uniform distribution of the second kind. Then,
	\begin{enumerate}
		\item[(i)]  The probability function of the marginal distribution of $\big(X_1, X_2,\ldots,X_r\big)$ for $1\leq r< k,$ is presented in the form:
		\begin{eqnarray}\label{i1s}
		P\big(X_1=x_1,X_2=x_2,\ldots,X_r=x_r\big)=\Phi(j,r) \frac{\left[\begin{array}{c}k-r+n-y_r \\n-y_r\end{array} \right]_{\mathcal{R}(p,q)}}{\left[\begin{array}{c}k+n \\n\end{array} \right]_{\mathcal{R}(p,q)}},
		\end{eqnarray}
		where  $\Phi(j,r)=\frac{\tau^{\sum_{j=1}^{r}(k-j+1)x_j}_2}{\tau^{\sum_{j=1}^{r}(k-j+1)x_j+2kn+ {k+1\choose 2}}_1},$  $x_j\in\{0,1,\ldots,n\}, j\in\{1,2,\ldots,r\},$ with $\sum_{j=1}^{r}x_j\leq n, y_r=\sum_{j=1}^{r}x_j.$
		\item[(ii)]The conditional  probability distribution of the vector $\big(X_{r+1}, X_{r+2},\ldots,X_{m}\big),$ given that $\big(X_1, X_2,\ldots,X_r\big)=\big(x_1, x_2,\ldots,x_r\big)$ for $1\leq r<m\leq k,$ turns out to be of the form:
		\begin{align}\label{ii1s}
		P\big(X_{r+1}=x_{r+1},\ldots,X_m=x_m|X_1=x_1,&\ldots,X_r=x_r\big)= \Phi(r,m)\nonumber\\&\times\frac{\left[\begin{array}{c}k-m+n-y_m \\n-y_m\end{array} \right]_{\mathcal{R}(p,q)}}{\left[\begin{array}{c}k-r+n-y_r \\n-y_r\end{array} \right]_{\mathcal{R}(p,q)}},
		\end{align}
		where $\Phi(r,m)=\frac{\tau^{\sum_{j=r+1}^{m}(k-j+1)x_j}_2}{\tau^{\sum_{j=r+1}^{m}(k-j+1)x_j+2kn+ {k+1\choose 2}}_1},$ $x_j\in\{0,1,\ldots,n-y_r\},$ and  $j\in\{r+1,r+2,\ldots,m\},$ with $\sum_{j=r+1}^{m}x_j\leq n-y_r, y_r=\sum_{i=1}^{r}x_i.$
	\end{enumerate}
\end{theorem} 
{\it Proof.} 
\begin{enumerate}
	\item[(i)]Summing the probability function of the multivariate discrete $\mathcal{R}(p,q)$-uniform
	distribution of the second kind, for $x_j\in\{0,1,\ldots,n-y_r\},$ and $j\in\{r+1,r+2,\ldots,k\},\, \sum_{j=r+1}^{k}x_j\leq n-y_r,$ we obtain the expression
	 for the marginal probability function of $(X_1, X_2, \ldots, X_r ):$ 
	\begin{align*}
	P\big(X_1=x_1,\ldots&,X_r=x_r\big)=\Phi(j,r)\,\frac{\sum \Phi(j,k-r)}{\left[\begin{array}{c} k+n  \\ n\end{array} \right]_{\mathcal{R}(p,q)}},
	\end{align*}
	where $\Phi(j,k-r)=\tau^{-\sum_{j=1}^{k-r}(k-r-j+1)x_{r+j}+2kn+ {k+1\choose 2}}_1\tau^{\sum_{j=1}^{k-r}(k-r-j+1)x_{r+j}}_2,$ and the summation is over  $x_{r+j}\in\{0,1,\ldots,n-y_r\}$ and $j\in\{1,2,\ldots,k-r\},$ with $\sum_{j=r+1}^{k}x_j\leq n-y_r.$ Then, the multiple sum, by using \eqref{hs2}, equals
	\begin{eqnarray*}
		\sum \tau^{-\sum_{j=1}^{k-r}(k-r-j+1)x_{r+j}}_1\tau^{\sum_{j=1}^{k-r}(k-r-j+1)x_{r+j}}_2=\left[\begin{array}{c} k-r+n-y_r  \\ n-y_r\end{array} \right]_{\mathcal{R}(p,q)},
	\end{eqnarray*}
	where the summation is over  $x_{r+j}\in\{0,1,\ldots,n-y_r\},$ and $j\in\{1,2,\ldots,k-r\},$ with $\sum_{j=r+1}^{k}x_j\leq n-y_r;$ the last expression of probability function then reduces to \eqref{i1s}.
	\item[(ii)]The conditional probability of $(X_{r+1}, X_{r+2}, \ldots , X_m),$ given that
	$(X_1, X_2, \ldots , X_r ) = (x_1, x_2, \ldots, x_r ),$ is given by
	\begin{align*}
	P\big(X_{r+1}=x_{r+1},\ldots, X_m=x_m|X_1&=x_1,\ldots X_{r}=x_{r}\big)\\&=\frac{P\big(X_1=x_1, X_2=x_2\ldots X_m=x_m\big)}{P\big(X_1=x_1, X_2=x_2\ldots X_{r}=x_{r}\big)}.
	\end{align*}
	Then, using the result  \eqref{i1s}, 
	we conclude that 
	\begin{align*}
	P\big(X_{r+1}=x_{r+1},\ldots, X_m=x_m|X_1=x_1,\ldots &X_{r}=x_{r}\big)=\Phi(r,m)\\&\times\frac{\left[\begin{array}{c} k-m+n-y_m  \\ n-y_m\end{array} \right]_{\mathcal{R}(p,q)}}{\left[\begin{array}{c} k-r+n-y_r  \\ n-y_r\end{array} \right]_{\mathcal{R}(p,q)}},
	\end{align*}
	where $x_j\in\{0,1,\ldots,n-y_r\},\,j\in\{r+1,r+2,\ldots,m\},\, \sum_{j=r+1}^{m}x_j\leq n-y_r$ and $y_j=\sum_{i=1}^{j}x_i.$
\end{enumerate}

We suppose that the random vector $(X_1, X_2,\ldots, X_k )$ satisfies a multivariate
discrete $\mathcal{R}(p,q)$-uniform distribution of the second kind and we consider the random variables 
\begin{eqnarray}
Y_j=\sum_{i=s_{j-1}+1}^{s_j} X_i=\sum_{i=1}^{m_j}X_{s_{j-1}}+i,\quad j\in\{1,2,\ldots,r\},
\end{eqnarray}
where $m_i,i\in\{1,2,\ldots,r\}$ are positive integers and $s_j=\sum_{i=1}^{j}m_j, j\in\{1,2,\ldots,r\},$ with $s_r=k,$ and $s_0=0.$ Then, 
the interesting probabilistic behaviour of groups of successive urns (energy levels)
is determined in the following theorem.
\begin{theorem} The probability function of
	\begin{enumerate}
		\item[(i)] 
		the distribution of the random vector $(Y_1, Y_2, \ldots, Y_r )$ is given by:
		\begin{align}\label{ia}
			P\big(Y_1=y_1,\ldots,Y_r=y_r\big)&=\tau^{\sum_{j=1}^{r}(n-z_j-s_j)(m_j-1)}_1\tau^{\sum_{j=1}^{r}(k-s_j+1)y_j}_2\nonumber\\&\times\frac{\prod_{j=1}^{r}\left[\begin{array}{c} m_j + y_j-1 \\ y_j\end{array} \right]_{\mathcal{R}(p,q)}}{\left[\begin{array}{c} k+n  \\ n\end{array} \right]_{\mathcal{R}(p,q)}},
		\end{align}
		where $y_j\in\{0,1,\ldots,n\},\,j\in\{1,2,\ldots,r\},\,\sum_{j=1}^{r}y_j\leq n.$
		\item[(ii)] the marginal distribution of the random vector $(Y_1, Y_2,\ldots , Y_{\nu} ),$ for $1\leq \nu < r,$ is given by:
		\begin{align}\label{iia}
			P\big(Y_1=y_1&,Y_2=y_2,\ldots,Y_{\nu}=y_{\nu}\big)=\tau^{\sum_{j=1}^{\nu}(n-z_j-s_j)(m_j-1)}_1\tau^{\sum_{j=1}^{\nu}(k-s_j+1)y_j}_2\nonumber\\&\times\frac{\prod_{j=1}^{\nu}\left[\begin{array}{c} m_j+y_j-1  \\ y_j\end{array} \right]_{\mathcal{R}(p,q)}\left[\begin{array}{c} k-s_{\nu}+n-z_{\nu}  \\ n-z_{\nu}\end{array} \right]_{\mathcal{R}(p,q)} }{\left[\begin{array}{c} k+n  \\ n\end{array} \right]_{\mathcal{R}(p,q)}},
		\end{align}
		where $y_j\in\{0,1,\ldots,n\},\, z_{\nu}=\sum_{i=1}^{\nu}y_i,\,j\in\{1,2,\ldots,\nu\},\,\sum_{j=1}^{\nu}y_j\leq n.$
		\item[(iii)]  the conditional distribution of the random vector $(Y_{\nu+1}, Y_{\nu+2},\ldots , Y_{\nu} ),$ given
		that $(Y_1, Y_2,\ldots, Y_{\nu} ) = (y_1, y_2, \ldots, y_{\nu} ),$ for $1\leq \nu <k\leq r,$ is given by
		\begin{align}\label{iiia}
			P\big(Y_{\nu+1}&=y_{\nu+1},\ldots,Y_{k}=y_{k}|Y_1=y_1,Y_2=y_2,\ldots,Y_{\nu}=y_{\nu}\big)=\nonumber\\&\times\tau^{\sum_{j=\nu+1}^{k}(n-z_j-s_j)(m_j-1)}_1\tau^{\sum_{j=\nu+1}^{k}(k-s_j+1)y_j}_2\nonumber\\&\times\frac{\prod_{j=\nu+1}^{k}\left[\begin{array}{c} m_j +y_j-1 \\ y_j\end{array} \right]_{\mathcal{R}(p,q)}\left[\begin{array}{c} k-s_{k}+n-z_{k}  \\ n-z_{k}\end{array} \right]_{\mathcal{R}(p,q)} }{\left[\begin{array}{c} k-s_{\nu}+n-z_{\nu}  \\ n-z_{\nu}\end{array} \right]_{\mathcal{R}(p,q)}},
		\end{align}
		where $y_j\in\{0,1,\ldots,n-z_{\nu}\},\,j\in\{\nu+1,\nu+2,\ldots,k\},\,\sum_{j=\nu+1}^{r}y_j\leq n,\,$ and $z_j=\sum_{i=1}^{j}y_i.$
	\end{enumerate}
\end{theorem}
{\it Proof.}
	\begin{enumerate}
		\item[(i)] The probability function of the random vector $(Y_1, Y_2, \ldots, Y_r )$ is deduced
		from the probability function
		\begin{eqnarray*}
			P\big(X_1=x_1,\ldots,X_k=x_k\big)=\Phi(j,k)\bigg/\left[\begin{array}{c} k+n  \\ n\end{array} \right]_{\mathcal{R}(p,q)},
		\end{eqnarray*}
		with $\Phi(j,k)=\tau^{-\sum_{j=1}^{k}(k-j+1)x_j+2kn+ {k+1\choose 2}}_1\tau^{\sum_{j=1}^{k}(k-j+1)x_j}_2,$	by inserting into it the $r$ new variables $(y_1, y_2, \ldots, y_r )$ and summing the resulting
		expression over all the remaining $k-r$ old variables. 
		The sum in the exponent of $\tau_1$ and $\tau_2$ can be expressed as:
		\begin{align*}
			\sum_{j=1}^{r}(k-j+1)x_j&=\sum_{j=1}^{r}\sum_{i=s_{j-1}+1}^{s_j}(k-i+1)x_i\\&=\sum_{j=1}^{r}\sum_{i=s_{j-1}+1}^{s_{j-1}+m_j}(k-i+1)x_i
		\end{align*}
		and
		\begin{align*}
			-\sum_{j=1}^{r}(k-j+1)x_j+2kn+ {k+1\choose 2}&=-\sum_{j=1}^{r}\sum_{i=s_{j-1}+1}^{s_{j-1}+m_j}(k-i+1)x_i\\&+2kn+ {k+1\choose 2}
		\end{align*}
		Besides, replacing in the last inner sum  the variable $i$ by $s_{j-1}+i$ and inserting
		into the resulting expression the variables $(y_1, y_2, \ldots, y_r ),$ we get
		\begin{align*}
			\sum_{j=1}^{k}(k-j+1)x_j&=\sum_{j=1}^{r}\sum_{i=1}^{m_j}(k-s_{j-1}-i+1)x_{s_{j-1}+i}\\&=\sum_{j=1}^{r}(k-s_j+1)\sum_{i=1}^{m_j}x_{s_{j-1}+i}+\sum_{j=1}^{r}\sum_{i=1}^{m_j-1}(m_j-i)x_{s_{j-1}+i}\\&=\sum_{j=1}^{r}(k-s_j+1)y_j+\sum_{j=1}^{r}\sum_{i=1}^{m_j-1}(m_j-i)x_{s_{j-1}+i}
		\end{align*}
		and 
		\begin{align*}
			-\sum_{j=1}^{r}(k-j+1)x_j-2kn+ {k+1\choose 2}&=-\sum_{j=1}^{r}(k-s_j+1)y_j+{k+1\choose 2}\\& +2kn+\sum_{j=1}^{r}\sum_{i=1}^{m_j-1}(m_j-i)x_{s_{j-1}+i}
		\end{align*}
		Thus, the probability function of the random vector $(Y_1, Y_2, \ldots, Y_r )$ is given by
		\begin{align*}
			P\big(Y_1=y_1,\ldots,Y_r=y_r\big)&=\frac{\psi(n,k)}{\left[\begin{array}{c} k+n  \\ n\end{array} \right]_{\mathcal{R}(p,q)}}\sum\tau^{\sum_{j=1}^{r}\sum_{i=1}^{m_j-1}(m_j-i)x_{s_{j-1}+i}}_2\\&\times \tau^{\sum_{j=1}^{r}\sum_{i=1}^{m_j-1}(m_j-i)x_{s_{j-1}+i}}_1\\&=\frac{\psi(n,k)}{\left[\begin{array}{c} k+n  \\ n\end{array} \right]_{\mathcal{R}(p,q)}}\prod_{j=1}^{r}\sum\tau^{\sum_{i=1}^{m_j-1}(m_j-i)x_{s_{j-1}+i}}_2\\&\times \tau^{-\sum_{i=1}^{m_j-1}(m_j-i)x_{s_{j-1}+i}}_1,
		\end{align*}
		where $\psi(n,k)=\tau^{\sum_{j=1}^{r}(k-s_j+1)y_j-2kn+ {k+1\choose 2}}_1\tau^{\sum_{j=1}^{r}(k-s_j+1)y_j}_2$ and the summation, in the last sum, is extended over all $x_{s_{j-1}+i}\in\{0,1,\ldots,y_j\},$ for $i\in\{0,1,\ldots,m_{j}-1\},$ with $\sum_{i=1}^{m_j-1}x_{s_{j-1}+i}\leq y_j.$ In addition to these values, the summation
		in the first sum is extended to all $j\in\{1,2,\ldots r\}.$ Moreover, by using \eqref{hs2}
		\begin{eqnarray*}
			\sum\frac{\tau^{-\sum_{i=1}^{m_j-1}(m_j-i)x_{s_{j-1}+i}}_1}{\tau^{\sum_{i=1}^{1-m_j}(m_j-i)x_{s_{j-1}+i}}_2}=\left[\begin{array}{c} m_j+y_j-1  \\ y_j\end{array} \right]_{\mathcal{R}(p,q)},
		\end{eqnarray*}
		where the sommation is over $x_{s_{j-1}+i}\in\{0,1\},\,i\in\{0,1,\ldots,m_{j}-1\}$ with $x_{s_{j-1}+1}+x_{s_{j-1}+2}+\ldots+ x_{s_{j-1}+m_j-1}\leq y_j,$ the expression \eqref{iia} is readily obtained. 
		\item[(ii)]Summing the probability function of the random vector $(Y_1, Y_2, \ldots, Y_r ),$ for $y_j\in\{0,1,\ldots,m_j\},\,j\in\{\nu+1,\nu+2,\ldots,r\},$ with $\sum_{j=\nu+1}^{r}y_j\leq n-z_r,$
		\begin{align*}
			P\big(Y_1=y_1,\ldots,Y_{\nu}=y_{\nu}\big)&=\tau^{\sum_{j=\nu+1}^{\nu}(n-z_j-s_j)(m_j-1)}_1\tau^{\sum_{j=1}^{\nu}(k-s_j+1)y_j}_2\\&\times\frac{\prod_{j=1}^{\nu}\left[\begin{array}{c} m_j+y_j-1  \\ y_j\end{array} \right]_{\mathcal{R}(p,q)}}{\left[\begin{array}{c} k+1  \\ n\end{array} \right]_{\mathcal{R}(p,q)}}\tau^{\sum_{j=\nu+1}^{\nu}(n-z_j-s_j)(m_j-1)}_1\\&\times\tau^{\sum_{j=\nu+1}^{r}(k-s_j+1)y_j}_2\prod_{j=\nu+1}^{r}\left[\begin{array}{c} m_j+y_j-1  \\ y_j\end{array} \right]_{\mathcal{R}(p,q)},
		\end{align*}
		where $y_{\nu+j}\in\{0,1,\ldots,n-z_{\nu}\},\,j\in\{1,2,\ldots,r-\nu\},\,y_{\nu+1}+\ldots+y_r\leq n-z_{\nu}$ and using \eqref{hsb},
		\begin{align*}
			\left[\begin{array}{c} k-s_{\nu}+n-z_{\nu}  \\ n-z_{\nu}\end{array} \right]_{\mathcal{R}(p,q)}&=\sum \tau^{\sum_{j=\nu+1}^{r}(n-z_j-s_j)(m_j-1)}_1\tau^{\sum_{j=\nu+1}^{r}(k-s_j+1))y_j}_2\\&\times\prod_{j=\nu+1}^{r}\left[\begin{array}{c} m_j+y_j-1  \\ y_j\end{array} \right]_{\mathcal{R}(p,q)} ,
		\end{align*}
		with $y_{\nu+j}\in\{0,1,\ldots,m_j\},\,j\in\{1,2,\ldots,r-\nu\},\,y_{\nu+1}+\ldots+y_r\leq n-z_{\nu},$ the probability function \eqref{iia} is obtained.
		\item[(iii)]The conditional probability  of $(Y_{\nu+1}, Y_{\nu+2},\ldots , Y_r ),$ given that
		$(Y_1, Y_2, \ldots, Y_r ) = (y_1, y_2,\ldots, y_r ),$ is provided by
		\begin{align*}
			P\big(Y_{\nu+1}=y_{\nu+1},\ldots, Y_r=y_r|Y_1&=y_1,\ldots Y_{\nu}=y_{\nu}\big)\\&=\frac{P\big(Y_1=y_1, Y_2=y_2\ldots Y_r=y_r\big)}{P\big(Y_1=y_1, Y_2=y_2\ldots Y_{\nu}=y_{\nu}\big)}.
		\end{align*}
		Then, by using the relations \eqref{ia} and \eqref{iia}, the formula is readily deduced. $\cqfd$
	\end{enumerate}
\subsection{$\mathcal{R}(p,q)$-bivariate uniform distribution of the second kind}
To investigate the $\mathcal{R}(p,q)$-factorial moments of the multivariate discrete  $\mathcal{R}(p,q)$-uniform
distribution of the second kind, and particularly the $\mathcal{R}(p,q)$-means, $\mathcal{R}(p,q)$-variance, and $\mathcal{R}(p,q)$-covariance,
we can restricted the computation  to the marginal distribution of the random
vector $(X_1, X_2).$ Taking $r=2,$ in the relation \eqref{i1s}, we obtain the probability function :
\begin{eqnarray}\label{rpqbusk}
P\big(X_1=x_1,X_2=x_2\big)=\Phi(j,2)\frac{\left[\begin{array}{c} k-2+n-x_1-x_2  \\ n-x_1-x_2\end{array} \right]_{\mathcal{R}(p,q)}}{\left[\begin{array}{c} k+n  \\ n\end{array} \right]_{\mathcal{R}(p,q)}},
\end{eqnarray}
where $\Phi(j,2)=\tau^{-kx_1-(k-1)x_2+2kn+ {k+1\choose 2}}_1\tau^{kx_1+(k-1)x_2}_2$ $x_j\in\{0,1,\ldots,n\},j\in\{1,2\},$ and $x_1+x_2\leq n.$ The $\mathcal{R}(p,q)$-factorial moments
of the random vector $(X_1, X_2)$ are derived in the following theorem.
\begin{theorem}
	Asumme that the random vector $(X_1, X_2)$ obeys the bivariate $\mathcal{R}(p,q)$-uniform distribution with probability function \eqref{rpqbusk}. Then,
	\begin{eqnarray}\label{rpqbuska}
	E\big([X_1]_{i_1,\mathcal{R}(p,q)}\big)=\frac{\tau^{2kn+ {k+1\choose 2}}_1\tau^{ki_1}_2\,[n]_{i_1,\mathcal{R}(p,q)}[i_1]_{\mathcal{R}(p,q)}!}{\tau^{ki_1}_1[k+i_1]_{i_1,\mathcal{R}(p,q)}},\quad i_1\in\{1,2,\ldots,n\},
	\end{eqnarray} 
	and
	\begin{align*}
		V\big([X_1]_{\mathcal{R}(p,q)}\big)&=\frac{\tau^{2k+1}_2\,[n]_{2,\mathcal{R}(p,q)}[2]_{\mathcal{R}(p,q)}!}{[k+2]_{2,\mathcal{R}(p,q)}}+\frac{\tau^{1-x_1}_1\tau^{k}_2\,[n]_{\mathcal{R}(p,q)}}{[k+1]_{\mathcal{R}(p,q)}}\\&-\frac{\tau^{2k}_2\,[n]^2_{\mathcal{R}(p,q)}}{[k+1]^2_{\mathcal{R}(p,q)}}.
	\end{align*}
Moreover, 
\begin{equation*}
E\big(\tau^{-i_2\,X_1}_1\tau^{i_2\,X_1}_2[X_2]_{i_2,\mathcal{R}(p,q)}\big)=\frac{\tau^{(1-k)i_2}_1\tau^{(k-1)i_2}_2[i_2]_{\mathcal{R}(p,q)}![n]_{i_2,\mathcal{R}(p,q)}}{\tau^{-2kn- {k+1\choose 2}}_1[k+i_2]_{i_2,\mathcal{R}(p,q)}}, 
\end{equation*}
with $i_2\in\{0,1,\ldots,n\}$ and
\begin{small}
\begin{eqnarray}
Cov\big([X_1]_{\mathcal{R}(p,q)},\tau^{-X_1}_1\tau^{X_1}_2[X_2]_{\mathcal{R}(p,q)}\big)=\frac{\tau^{2k-1}_2[n]_{\mathcal{R}(p,q)}\nabla(p,q)}{\tau^{2k(1-n)-{k+1\choose 2}}_2[k+1]^2_{\mathcal{R}(p,q)}[k+2]_{\mathcal{R}(p,q)}},
\end{eqnarray}
\end{small}
where $$\nabla(p,q)=\tau_2[n-1]_{\mathcal{R}(p,q)}[k+1]_{\mathcal{R}(p,q)}-\tau_1[n]_{\mathcal{R}(p,q)}[k+2]_{\mathcal{R}(p,q)}.$$
\end{theorem}
{\it Proof.} 
	The marginal probability function of $X_1$ is
	\begin{eqnarray}
	P\big(X_1=x_1\big)=\tau^{-kx_1+2kn+ {k+1\choose 2}}_1\tau^{kx_1}_2\frac{\left[\begin{array}{c} k-1+n-x_1  \\ n-x_1\end{array} \right]_{\mathcal{R}(p,q)}}{\left[\begin{array}{c} k+n  \\ n\end{array} \right]_{\mathcal{R}(p,q)}}, \quad x_1\in\mathbb{N}
	\end{eqnarray}
	and its $\mathcal{R}(p,q)$-factorial moments are given by:
	\begin{align*}
	E\big([X_1]_{i_1,\mathcal{R}(p,q)}\big)&=\sum_{x_1=i_1}^{n}\tau^{-kx_1+2kn+ {k+1\choose 2}}_1\tau^{kx_1}_2[x_1]_{i_1,\mathcal{R}(p,q)}\\&\times\frac{\left[\begin{array}{c} k-1+n-x_1  \\ n-x_1\end{array} \right]_{\mathcal{R}(p,q)}}{\left[\begin{array}{c} k+n  \\ n\end{array} \right]_{\mathcal{R}(p,q)}}\\&=\tau^{2kn+ {k+1\choose 2}}_1[i_1]_{\mathcal{R}(p,q)}!\sum_{x_1=i_1}^{n}\tau^{-kx_1}_1\tau^{kx_1}_2\\&\times\left[\begin{array}{c} x_1  \\ x_1-i_1\end{array} \right]_{\mathcal{R}(p,q)}\frac{\left[\begin{array}{c} k-1+n-x_1  \\ n-x_1\end{array} \right]_{\mathcal{R}(p,q)}}{\left[\begin{array}{c} k+n  \\ n\end{array} \right]_{\mathcal{R}(p,q)}}.
	\end{align*}
	Putting $r=x_1-i_1$ and from the $\mathcal{R}(p,q)$-Cauchy's formula,
	\begin{small}
	\begin{eqnarray*}
	\sum_{r=0}^{n}\tau^{(m-k)r}_1\tau^{(k-m)r}_2\left[\begin{array}{c} m+r  \\ r\end{array} \right]_{\mathcal{R}(p,q)}\left[\begin{array}{c} k-m+n-r-1  \\ n-r\end{array} \right]_{\mathcal{R}(p,q)}=\left[\begin{array}{c} k+n  \\ n\end{array} \right]_{\mathcal{R}(p,q)},
	\end{eqnarray*}
\end{small}
	with $m=i_1,$ and $k$ and $n$ replaced by $k+i_1$ and $n-i_1,$ respectively, the relation \eqref{rpqbuska} follows. Moreover, the $\mathcal{R}(p,q)$-variance of the random variable is given by
	\begin{equation*}
	 V\big([X_1]_{\mathcal{R}(p,q)}\big)=\tau_2\,{\bf E}\big({[X_1]_{2,\mathcal{R}(p,q)}}\big) +\tau^{X-1}_1\, E\big([X_1]_{\mathcal{R}(p,q)}\big)- \big[ E\big([X_1]_{\mathcal{R}(p,q)}\big)\big]^2.
	\end{equation*}
	By using the relation \eqref{rpqbuska}, we have:
	\begin{align*}
	V\big([X_1]_{\mathcal{R}(p,q)}\big)&=\frac{\tau^{2k+1}_2\,[n]_{2,\mathcal{R}(p,q)}[2]_{\mathcal{R}(p,q)}!}{[k+2]_{2,\mathcal{R}(p,q)}}+\frac{\tau^{1-x_1}_1\tau^{k}_2\,[n]_{\mathcal{R}(p,q)}}{[k+1]_{\mathcal{R}(p,q)}}\\&-\frac{\tau^{2k}_2\,[n]^2_{\mathcal{R}(p,q)}}{[k+1]^2_{\mathcal{R}(p,q)}}.
	\end{align*}
Besides, the expected value of $\tau^{i_2\,X_1}_2[X_2]_{i_2,\mathcal{R}(p,q)}$ may be computed by using the relation:
\begin{eqnarray*}
E\big(\tau^{-i_2\,X_1}_1\tau^{i_2\,X_1}_2[X_2]_{i_2,\mathcal{R}(p,q)}\big)=E\big[E\big(\tau^{-i_2\,X_1}_1\tau^{i_2\,X_1}_2[X_2]_{i_2,\mathcal{R}(p,q)}|X_1\big)\big].
\end{eqnarray*}
The conditional probability function of the random variable $X_2,$ given that $X_1=x_1,$ is given by 
\begin{eqnarray}
P\big(X_2=x_2|X_1=x_1\big)&=&\tau^{-(k-1)x_2+2kn+ {k+1\choose 2}}_1\tau^{(k-1)x_2}_2\nonumber\\&\times&\frac{\left[\begin{array}{c} k-2+n-x_1-x_2  \\ n-x_1-x_2\end{array} \right]_{\mathcal{R}(p,q)}}{\left[\begin{array}{c} k-1+n-x_1-x_2  \\ n-x_1\end{array} \right]_{\mathcal{R}(p,q)}},
\end{eqnarray}
where $x_2\in\{0,1,\ldots,n-x_1\}.$ Replacing $k$ and $n$ by $k-1$ and $n-x_1$ in the above relation, respectively, we obtain the probability function of the random varaible $X_1.$ Then, 
\begin{eqnarray*}
E\big([X_2]_{i_2,\mathcal{R}(p,q)}|X_1=x_1\big)=\frac{\tau^{2kn+ {k+1\choose 2}}_1\tau^{(k-1)i_2}_2\,[n-x_1]_{i_2,\mathcal{R}(p,q)}[i_2]_{\mathcal{R}(p,q)}!}{\tau^{(k-1)i_2}_1[k+i_2-1]_{i_2,\mathcal{R}(p,q)}}.
\end{eqnarray*}
Moreover, the expected value of $\tau^{-i_2\,X_1}_1\tau^{i_2\,X_1}_2[n-X_1]_{i_2,\mathcal{R}(p,q)}$ is given by:
\begin{align*}
E\big(\tau^{-i_2\,X_1}_1\tau^{i_2\,X_1}_2[n-&X_1]_{i_2,\mathcal{R}(p,q)}\big)=\sum_{x_1=0}^{n-i_2}\tau^{-(k+i_2)x_1+2kn+ {k+1\choose 2}}_1\tau^{(k+i_2)x_1}_2\\&\times[n-x_1]_{i_2,\mathcal{R}(p,q)}\frac{\left[\begin{array}{c} k-1+n-x_1  \\ n-x_1\end{array} \right]_{\mathcal{R}(p,q)}}{\left[\begin{array}{c} k+n  \\ n\end{array} \right]_{\mathcal{R}(p,q)}}\\
&=[k+i_2-1]_{i_2,\mathcal{R}(p,q)}\tau^{2kn+ {k+1\choose 2}}_1\\&\times\sum_{x_1=0}^{n-i_2}\tau^{-(k+i_2)x_1}_1\tau^{(k+i_2)x_1}_2\frac{\left[\begin{array}{c} k-1+n-x_1  \\ n-x_1\end{array} \right]_{\mathcal{R}(p,q)}}{\left[\begin{array}{c} k+n  \\ n\end{array} \right]_{\mathcal{R}(p,q)}}.
\end{align*}
By using the $\mathcal{R}(p,q)$-Cauchy's formula with $m=0,$ and $k$ and $n$ replaced by $k+i_1$ and $n-i_1,$ respectively, we obtain:
\begin{eqnarray*}
E\big(\tau^{-i_2\,X_1}_1\tau^{i_2\,X_1}_2[n-X_1]_{i_2,\mathcal{R}(p,q)}\big)=\frac{[n]_{i_2,\mathcal{R}(p,q)}[k+i_2-1]_{i_2,\mathcal{R}(p,q)}}{\tau^{-2kn- {k+1\choose 2}}_1[k+i_2]_{i_2,\mathcal{R}(p,q)}}
\end{eqnarray*}
and 
\begin{align*}
	E\big(\tau^{-i_2\,X_1}_1\tau^{i_2\,X_1}_2[X_2]_{i_2,\mathcal{R}(p,q)}\big)&=E\big[E\big(\tau^{-i_2\,X_1}_1\tau^{i_2\,X_1}_2[X_2]_{i_2,\mathcal{R}(p,q)}|X_1\big)\big]\\
	&=\frac{\tau^{(1-k)i_2}_1\tau^{(k-1)i_2}_2[i_2]_{\mathcal{R}(p,q)}!}{[k+i_2-1]_{i_2,\mathcal{R}(p,q)}}\\&\times E\big(\tau^{-i_2\,X_1}_1\tau^{i_2\,X_1}_2[n-X_1]_{i_2,\mathcal{R}(p,q)}\big)\\&=\frac{\tau^{(1-k)i_2}_1\tau^{(k-1)i_2}_2[i_2]_{\mathcal{R}(p,q)}!}{[k+i_2-1]_{i_2,\mathcal{R}(p,q)}}\\&\times\frac{[n]_{i_2,\mathcal{R}(p,q)}[k+i_2-1]_{i_2,\mathcal{R}(p,q)}}{\tau^{-2kn- {k+1\choose 2}}_1[k+i_2]_{i_2,\mathcal{R}(p,q)}}\\&=\frac{\tau^{(1-k)i_2}_1\tau^{(k-1)i_2}_2[i_2]_{\mathcal{R}(p,q)}![n]_{i_2,\mathcal{R}(p,q)}}{\tau^{-2kn- {k+1\choose 2}}_1[k+i_2]_{i_2,\mathcal{R}(p,q)}}.
\end{align*}
By analogy, the expected value of the $\mathcal{R}(p,q)$-function $\tau^{-i_2X_1}_1\tau^{i_2X_1}_2[X_1]_{i_1,\mathcal{R}(p,q)}[X_2]_{i_2,\mathcal{R}(p,q)}$ can be computed from the relation
\begin{eqnarray*}
	E\big(\tau^{-i_2X_1}_1\tau^{i_2X_1}_2[X_1]_{i_1,\mathcal{R}(p,q)}[X_2]_{i_2,\mathcal{R}(p,q)}\big)=E\big[E\big(\tau^{-i_2X_1}_1\tau^{i_2X_1}_2[X_1]_{i_1,\mathcal{R}(p,q)}[X_2]_{i_2,\mathcal{R}(p,q)}|X_1\big)\big].
\end{eqnarray*}
Naturally, 
\begin{small}
\begin{align*}
E\bigg(\frac{\tau^{-i_2X_1}_1}{\tau^{i_2X_1}_2}[X_1]_{i_1,\mathcal{R}(p,q)}[n-&X_1]_{i_2,\mathcal{R}(p,q)}\bigg)=\sum_{x_1=i_1}^{n-i_2}\frac{\tau^{-(k+i_2)x_1+2kn+ {k+1\choose 2}}_1}{\tau^{-(k+i_2)x_1}_2}[x_1]_{i_1,\mathcal{R}(p,q)}\\&\times[n-x_1]_{i_2,\mathcal{R}(p,q)}\frac{\left[\begin{array}{c} k-1+n-x_1  \\ n-x_1\end{array} \right]_{\mathcal{R}(p,q)}}{\left[\begin{array}{c} k+n  \\ n\end{array} \right]_{\mathcal{R}(p,q)}}\\
&=[i_1]_{\mathcal{R}(p,q)}![k+i_2-1]_{i_2,\mathcal{R}(p,q)}\\&\times\sum_{x_1=i_1}^{n-i_2}\tau^{-(k+i_2)x_1+2kn+ {k+1\choose 2}}_1\tau^{(k+i_2)x_1}_2\\&\times\frac{\left[\begin{array}{c} x_1  \\ x_1-i_1\end{array} \right]_{\mathcal{R}(p,q)}\left[\begin{array}{c} k-1+n-x_1  \\ n-i_2-x_1\end{array} \right]_{\mathcal{R}(p,q)}}{\left[\begin{array}{c} k+n  \\ n\end{array} \right]_{\mathcal{R}(p,q)}}
\end{align*}
\end{small}
and from the $\mathcal{R}(p,q)$-Cauchy's formula with $m=i_1,$ and  $k$ and $n$ replaced by $k+i_1$ and $n-i_1,$ respectively, the above relation is reduced to
\begin{align*}
E\big(\tau^{-i_2X_1}_1\tau^{i_2\,X_1}_2[X_1]_{i_1,\mathcal{R}(p,q)}[n-X_1]_{i_2,\mathcal{R}(p,q)}\big)
&=\frac{\tau^{-(k+i_2)i_1}_1\tau^{(k+i_2)i_1}_2[i_1]_{\mathcal{R}(p,q)}!}{\tau^{-2kn-{k+1\choose 2}}_1}\\&\times\frac{[n]_{i_1+i_2,\mathcal{R}(p,q)}[k+i_2-1]_{i_2,\mathcal{R}(p,q)}}{[k+i_1+i_2]_{i_1+i_2,\mathcal{R}(p,q)}}.
\end{align*}
Thus, 
\begin{small}
\begin{align*}
	E\bigg(\frac{\tau^{-i_2X_1}_1}{\tau^{-i_2X_1}_2}[X_1]_{i_1,\mathcal{R}(p,q)}[X_2]_{i_2,\mathcal{R}(p,q)}\bigg)&=\frac{\tau^{k(i_1+i_2)+(i_1-1)i_2}_2[i_2]_{\mathcal{R}(p,q)}!}{\tau^{k(i_1+i_2)+(i_1-1)i_2+2kn+{k+1\choose 2}}_1}\\&\times \frac{[i_1]_{\mathcal{R}(p,q)}![n]_{i_1+i_2,\mathcal{R}(p,q)}}{[k+i_1+i_2]_{i_1+i_2,\mathcal{R}(p,q)}}.
\end{align*}
\end{small}
The covariance of the random variables $[X_1]_{\mathcal{R}(p,q)}$ and $\tau^{-X_1}_1\tau^{X_1}_2[X_2]_{\mathcal{R}(p,q)}$ is given by
\begin{align*}
Cov\big([X_1]_{\mathcal{R}(p,q)},\tau^{-X_1}_1\tau^{X_1}_2[X_2]_{\mathcal{R}(p,q)}\big)&=E\big(\tau^{-X_1}_1\tau^{X_1}_2[X_1]_{\mathcal{R}(p,q)}[X_2]_{\mathcal{R}(p,q)}\big)\\&-E\big([X_1]_{\mathcal{R}(p,q)}\big)E\big(\tau^{-X_1}_1\tau^{X_1}_2[X_2]_{\mathcal{R}(p,q)}\big)\\&=\frac{\tau^{2k}_2[n]_{2,\mathcal{R}(p,q)}}{\tau^{2k(1-n)-{k+1\choose 2}}_1[k+2]_{2,\mathcal{R}(p,q)}}\\&-\frac{\tau^{1-2k}_1\tau^{2k-1}_2[n]^2_{\mathcal{R}(p,q)}}{\tau^{-2k-{k+1\choose 2}}_1[k+1]^2_{\mathcal{R}(p,q)}}.
\end{align*}
By computation, the result follows and the proof is achieved. $\cqfd$
\subsection{Particular cases of multivariate uniform distribution of second kind}
The probability function of the multivariate discrete $(p,q)$-uniform distribution of the second
kind, with parameters $n,$ $p$ and $q,$ is given by:
\begin{eqnarray*}
P\big(X_1=x_1, X_2=x_2,\ldots, X_k=x_k\big)=\Phi(n,k)\frac{1}{\left[\begin{array}{c}k+n \\n\end{array} \right]_{p^{-1},q^{-1}}},
\end{eqnarray*}
where $\Phi(n,k)=p^{-\sum_{j=1}^{k}(k-j+1)x_j+2kn+ {k+1\choose 2}}\,q^{\sum_{j=1}^{k}(k-j+1)x_j}.$ Moreover, we assume that the random vector $\big(X_1, X_2,\ldots,X_k\big)$ satisfies a multivariate
discrete $(p,q)$-uniform distribution of the second kind. Then, 
\begin{enumerate}
	\item[(i)] The probability function of the marginal distribution of $\big(X_1, X_2,\ldots,X_r\big)$ for $1\leq r< k,$ is presented in the form:
	\begin{eqnarray*}
	P\big(X_1=x_1,X_2=x_2,\ldots,X_r=x_r\big)= \frac{q^{\sum_{j=1}^{r}(k-j+1)x_j}\left[\begin{array}{c}k-r+n-y_r \\n-y_r\end{array} \right]_{p,q}}{p^{\sum_{j=1}^{r}(k-j+1)x_j-2kn- {k+1\choose 2}}\left[\begin{array}{c}k+n \\n\end{array} \right]_{p,q}},
	\end{eqnarray*}
	where  $x_j\in\{0,1,\ldots,n\}, j\in\{1,2,\ldots,r\},$ with $\sum_{j=1}^{r}x_j\leq n, y_r=\sum_{j=1}^{r}x_j.$
	\item[(ii)]The conditional  probability distribution of the vector $\big(X_{r+1}, X_{r+2},\ldots,X_{m}\big),$ given that $\big(X_1, X_2,\ldots,X_r\big)=\big(x_1, x_2,\ldots,x_r\big)$ for $1\leq r<m\leq k,$ is given by:
	\begin{align*}
		P\big(X_{r+1}=x_{r+1}&,\ldots,X_m=x_m|X_1=x_1,\ldots,X_r=x_r\big)=\\& \frac{q^{\sum_{j=r+1}^{m}(k-j+1)x_j}}{p^{\sum_{j=r+1}^{m}(k-j+1)x_j-2kn- {k+1\choose 2}}}\frac{\left[\begin{array}{c}k-m+n-y_m \\n-y_m\end{array} \right]_{p,q}}{\left[\begin{array}{c}k-r+n-y_r \\n-y_r\end{array} \right]_{p,q}},
	\end{align*}
	where  $x_j\in\{0,1,\ldots,n-y_r\}, j\in\{r+1,r+2,\ldots,m\},$ with $\sum_{j=r+1}^{m}x_j\leq n-y_r, y_r=\sum_{i=1}^{r}x_i.$
\end{enumerate}
Furthermore, 
		we suppose that the random vector $(X_1, X_2,\ldots, X_k )$ obeys a multivariate
		discrete $(p,q)$-uniform distribution of the second kind, and we consider the random variables 
		\begin{eqnarray*}
			Y_j=\sum_{i=s_{j-1}+1}^{s_j} X_i=\sum_{i=1}^{m_j}X_{s_{j-1}}+i,\quad j\in\{1,2,\ldots,r\},
		\end{eqnarray*}
		where $m_i,i\in\{1,2,\ldots,r\}$ are positive integers and $s_j=\sum_{i=1}^{j}m_j, j\in\{1,2,\ldots,r\},$ with $s_r=k,$ and $s_0=0.$ Then:
		\begin{enumerate}
			\item[(i)] 
			The probability function of
			the distribution of the  vector $(Y_1, Y_2, \ldots, Y_r )$ is given by:
			\begin{eqnarray*}
			P\big(Y_1=y_1,\ldots,Y_r=y_r\big)=\frac{p^{\sum_{j=1}^{r}(n-z_j-s_j)(m_j-1)}}{q^{-\sum_{j=1}^{r}(k-s_j+1)y_j}}\frac{\prod_{j=1}^{r}\left[\begin{array}{c} m_j + y_j-1 \\ y_j\end{array} \right]_{p,q}}{\left[\begin{array}{c} k+n  \\ n\end{array} \right]_{p,q}},
			\end{eqnarray*}
			where $y_j\in\{0,1,\ldots,n\},\,j\in\{1,2,\ldots,r\},\,\sum_{j=1}^{r}y_j\leq n.$
			\item[(ii)]The probability function of the marginal distribution of the vector $(Y_1, Y_2,\ldots , Y_{\nu} ),$ for $1\leq \nu < r,$ is yielded by:
			\begin{align*}
			P\big(Y_1=y_1,Y_2=y_2&,\ldots,Y_{\nu}=y_{\nu}\big)=p^{\sum_{j=1}^{\nu}(n-z_j-s_j)(m_j-1)}q^{\sum_{j=1}^{\nu}(k-s_j+1)y_j}\nonumber\\&\times\frac{\prod_{j=1}^{\nu}\left[\begin{array}{c} m_j+y_j-1  \\ y_j\end{array} \right]_{p,q}\left[\begin{array}{c} k-s_{\nu}+n-z_{\nu}  \\ n-z_{\nu}\end{array} \right]_{p,q} }{\left[\begin{array}{c} k+n  \\ n\end{array} \right]_{p,q}},
			\end{align*}
			where $y_j\in\{0,1,\ldots,n\},\, z_{\nu}=\sum_{i=1}^{\nu}y_i,\,j\in\{1,2,\ldots,\nu\},\,\sum_{j=1}^{\nu}y_j\leq n.$
			\item[(iii)]The probability  of  the conditional distribution of the vector $(Y_{\nu+1}, Y_{\nu+2},\ldots , Y_{\nu} ),$ given
			that $(Y_1, Y_2,\ldots, Y_{\nu} ) = (y_1, y_2, \ldots, y_{\nu} ),$ for $1\leq \nu <k\leq r,$ is given by:
			\begin{align*}
			P\big(Y_{\nu+1}&=y_{\nu+1},\ldots,Y_{k}=y_{k}|Y_1=y_1,Y_2=y_2,\ldots,Y_{\nu}=y_{\nu}\big)=\nonumber\\&\times p^{\sum_{j=\nu+1}^{k}(n-z_j-s_j)(m_j-1)}q^{\sum_{j=\nu+1}^{k}(k-s_j+1)y_j}\nonumber\\&\times\frac{\prod_{j=\nu+1}^{k}\left[\begin{array}{c} m_j +y_j-1 \\ y_j\end{array} \right]_{p,q}\left[\begin{array}{c} k-s_{k}+n-z_{k}  \\ n-z_{k}\end{array} \right]_{p,q} }{\left[\begin{array}{c} k-s_{\nu}+n-z_{\nu}  \\ n-z_{\nu}\end{array} \right]_{p,q}},
			\end{align*}
			where $y_j\in\{0,1,\ldots,n-z_{\nu}\},\,j\in\{\nu+1,\nu+2,\ldots,k\},\,\sum_{j=\nu+1}^{r}y_j\leq n,\,z_j=\sum_{i=1}^{j}y_i.$
		\end{enumerate}
	Particular cases are deduced as follows: 
	the $(p,q)$-bivariate uniform probability distribution 
	\begin{eqnarray*}
		P\big(X_1=x_1,X_2=x_2\big)=q^{kx_1+(k-1)x_2}\frac{\left[\begin{array}{c} k-2+n-x_1-x_2  \\ n-x_1-x_2\end{array} \right]_{p,q}}{\left[\begin{array}{c} k+n  \\ n\end{array} \right]_{p,q}},
	\end{eqnarray*}
	where $x_1\in\mathbb{N},x_2\in\mathbb{N},$and $x_1+x_2\leq n,$ with the properties:
	\begin{eqnarray*}
		E\big([X_1]_{i_1,p,q}\big)=\frac{q^{ki_1}\,[n]_{i_1,p,q}[i_1]_{p,q}!}{[k+i_1]_{i_1,p,q}},\quad i_1\in\{1,2,\ldots,n\},
	\end{eqnarray*} 
	and
	\begin{eqnarray*}
		V\big([X_1]_{p,q}\big)=\frac{q^{2k+1}\,[n]_{2,p,q}[2]_{p,q}!}{[k+2]_{2,p,q}}+\frac{p^{1-x_1}q^{k}\,[n]_{p,q}}{[k+1]_{p,q}}-\frac{q^{2k}\,[n]^2_{p,q}}{[k+1]^2_{p,q}}.
	\end{eqnarray*}
	Moreover, 
	\begin{equation*}
		E\big(\tau^{i_2\,X_1}_2[X_2]_{i_2,p,q}\big)=\frac{q^{(k-1)i_2}[i_2]_{p,q}![n]_{i_2,p,q}}{[k+i_2]_{i_2,p,q}},\quad i_2\in\{0,1,\ldots,n\}
	\end{equation*}
	and
	\begin{eqnarray*}
		Cov\big([X_1]_{p,q},q^{X_1}[X_2]_{p,q}\big)=\frac{\tau^{2k-1}_2[n]_{p,q}\nabla(p,q)}{[k+1]^2_{p,q}[k+2]_{p,q}},
	\end{eqnarray*}
	where $$\nabla(p,q)=q[n-1]_{p,q}[k+1]_{p,q}-[n]_{p,q}[k+2]_{p,q}.$$
\section{Concluding remarks}
The multivariate uniform probability distribution of the first and second kinds  from the  $\mathcal{R}(p,q)$-deformed quantum
algebras have been investigated. The corresponding  bivariate distributions,   properties such as ($\mathcal{R}(p,q)$-mean, variance and covariance), and 
particular cases have been deduced and discussed. 
\section*{Acknowledgments}
This research was partly supported by the SNF Grant No. IZSEZ0\_206010. The work of MNH  is supported by 
 the ICMPA-UNESCO Chair, which is in partnership with Daniel Iagolnitzer Foundation (DIF), France, and the Association pour la Promotion Scientifique de l'Afrique (APSA), supporting the development of mathematical physics in Africa.

\end{document}